\documentclass[12pt]{article}
\pdfoutput=1
\usepackage{graphicx}
\usepackage{epstopdf}
\usepackage{amsmath}
\usepackage{amsfonts}
\usepackage{amssymb}
\usepackage{color}
\usepackage{mathrsfs}

\newcommand{\be}{\begin{equation}}
\newcommand{\ee}{\end{equation}}
\newcommand{\bea}{\begin{eqnarray}}
\newcommand{\eea}{\end{eqnarray}}
\newcommand{\barr}{\begin{array}}
\newcommand{\earr}{\end{array}}
\newcommand{\ta}{\tilde a}
\newcommand{\vk}{\vec k}
\newcommand{\vkp}{\vec q}
\newcommand{\vkkp}{\vec k-\vec q}
\newcommand{\amp}{&\!\!}

\def\beq{\begin{equation}}
\def\eeq{\end{equation}}
\def\be{\begin{equation}}
\def\ee{\end{equation}}
\def\bea{\begin{eqnarray}}
\def\eea{\end{eqnarray}}
\def\d{{\partial}}

\def\ba{\begin{align}}
\def\ea{\end{align}}

\def\fig#1{fig.~{\ref{#1}}}
\def\Fig#1{Fig.~{\ref{#1}}}

\newcommand\myR{ {\vec{r}}}

\def\app#1{App.~{\ref{#1}}}
\def\csc{ {c^2_{\rm comb}}}

\begin{document}
\hspace{5.2in} \mbox{SU-ITP-12/17}


\setcounter{page}{1} \baselineskip=15.5pt \thispagestyle{empty}

\begin{flushright}
\end{flushright}

\begin{center}

\def\thefootnote{\fnsymbol{footnote}}

{\Large \bf The Effective Field Theory\\[0.3cm] of Cosmological Large Scale Structures}
\\[0.5cm]

{\large John Joseph M. Carrasco$^{1}$, Mark P. Hertzberg$^{1,2}$, Leonardo Senatore$^{1,2}$}
\\[0.5cm]

{\normalsize { \sl $^{1}$ Stanford Institute for Theoretical Physics and Department of Physics, \\Stanford University, Stanford, CA 94306}}\\
\vspace{.3cm}

{\normalsize { \sl $^{2}$ Kavli Institute for Particle Astrophysics and Cosmology, \\ Stanford University and SLAC, Menlo Park, CA 94025}}\\
\vspace{.3cm}

\end{center}

\vspace{.8cm}

\hrule \vspace{0.3cm}
{\small  \noindent \textbf{Abstract} \\[0.3cm]
Large scale structure surveys will likely become the next leading cosmological probe. In our universe, matter perturbations are large on short distances and small at long scales, i.e. strongly coupled in the UV and weakly coupled in the IR. To make precise analytical predictions on large scales, we develop an effective field theory formulated in terms of an IR effective fluid characterized by several parameters, such as speed of sound and viscosity. These parameters, determined by the UV physics described by the Boltzmann equation, are measured from $N$-body simulations. We find that the speed of sound of the effective fluid is $c_s^2 \approx 10^{-6}c^2$ and that the viscosity contributions are of the same order.
The fluid describes all the relevant physics at long scales $k$ and permits a manifestly convergent perturbative expansion in the size of the matter perturbations $\delta(k)$ for all the observables. As an example, we calculate the correction to the power spectrum at order $\delta(k)^4$.  The predictions of the effective field theory are found to be in much better agreement with observation than standard cosmological perturbation theory, already reaching percent precision at this order up to a relatively short scale~$k\simeq 0.24 h$~Mpc$^{-1}$.
\noindent 
}
 \vspace{0.3cm}
\hrule



\section{Introduction}

Large Scale Structure Surveys have the potential of becoming the leading cosmological observable in the next decade. They contain a  tremendous amount of cosmological information. If we were able to extract information from all the modes that go from the horizon scale $\sim 10^4$~Mpc to the non-linear scale $\sim 10$ Mpc, we would obtain about 
\be
\left(\frac{10^4}{10}\right)^3\sim 10^9
\ee
independent modes. The Planck satellite in comparison has about $(2 \times 10^3)^2\sim 10^6$ modes. Of course, accessing all this information is much harder than for the CMB, due to the short scale non-linearities. There are several aspects to this problem. The first problem is related to our currently limited understanding of the evolution of dark matter on large scales. Non-linear corrections are very important even on scales larger than $10$ Mpc, because modes of different wavelengths couple to each other. Understanding these corrections is a problem that affects all large scale structure observables. There are then two additional issues that affect most, but at least not all, observables. One is the fact that most dark matter is clumped in very non-linear structures (dark matter halos); and the other is the fact that what we often observe are galaxies, and not just dark matter halos, and not even dark matter long wavelength perturbations. The solution to these two last problems requires the correct understanding of the  so-called halo- and galaxy- biases. These two problems, while important and deeply interesting in their own right, are very astrophysical in nature, and we do not address them here. 

Instead here we try to address in a rigorous way the first problem, that is the prediction of the dark matter distribution on scales larger than the non-linear scale. The fact that the universe is characterized by two well separated scales, the Hubble scale, over which perturbations are linear, and the non-linear scale, which indeed characterizes the scale over which gravitational collapse overtakes the expansion of the universe, makes the problem amenable to an Effective Field Theory (EFT) treatment. An effective theory is a description of a  system that captures all the relevant degrees of freedom and describes all the relevant physics at a macroscopic scale of interest. The short distance (so called ultraviolet or `UV') physics is integrated out and affects the effective field theory only through various couplings in a perturbative expansion in the ratio of  microphysical UV scale/s to the macroscopic scale being probed. 
This technique has been systematically used in particle physics and condensed matter physics for many years, but has not been fully used in astrophysics and cosmology. An important early (and recent) application of these techniques in cosmology is the so-called Effective Field Theory of Inflation~\cite{Cheung:2007st}.  In a similar vein, understanding the large scale properties of the universe is very important, and is ready for a careful analysis.

Indeed the situation in the universe is very similar to what happens in the chiral Lagrangian that describes pion interactions in Particle Physics. At very low energies, pions are weakly interacting. These interactions and the size of the fluctuations grow with energy until we hit the Quantum ChromoDynamics (QCD) scale, $\sim4\pi F_\pi$, at which the pions become strongly coupled. The Chiral Lagrangian~\cite{Weinberg:1996kr} offers the correct effective theory allowing arbitrarily precise predictions, up to non-perturbative effects, at energies $E\ll 4\pi F_\pi$. In our universe, matter fluctuations are small at large distances and becomes larger and larger as we move up to the non-linear scale. Since the size of the non-linear terms, which are nothing but interactions, grows with the size of the fluctuations, we see that at long distances the universe should be described by some weakly coupled degree of freedom, that becomes more and more interacting as we move closer to the non-linear scale, at which point the fluctuations become strongly coupled. The coupling constant should indeed be represented by the ratio of the considered wavenumber $k$ over the wavenumber at the non-linear scale $k_{NL}$: $k/k_{NL}$. Notice that indeed the size of the density perturbations $\delta\rho/\rho$  on a scale $k$ scales as $(k/k_{NL})^2$. This scaling suggests the existence of an effective field theory that should allow us to describe with arbitrary precision the universe on scales $k\ll k_{NL}$,  very much as the Chiral Lagrangian represents the right effective field theory to describe pion dynamics to arbitrary precision.

Such an effective theory would have particularly relevant observational implications. Already now, large scale structure surveys such as BOSS or DES are measuring  the galaxy-galaxy correlation function, so called Baryon Acoustic Oscillations (BAO), at scales of order 100 Mpc. Next generation experiments such as LSST will measure this quantity at about percent precision. These observations contain huge amount of information on Dark Energy and on Inflation, through for example the non-Gaussianity of the primordial perturbations. The BAO scale is about one order of magnitude longer than the non-linear scale, where $\delta\rho/\rho\sim 10^{-2}$,  and therefore physics at this scale {\it must} be describable by rigorous perturbative methods.  The alternative is to rely on either  time consuming numerical simulations, or on analytical approaches that however are limited by some irreducible mistake that is hard to quantify precisely. In an ideal situation, numerical $N$-body simulations should be  quickly done only at {\it small} scales, to describe phenomena affected by gravitational collapse, rather than running large simulations to describe weakly coupled physics. This has been indeed recently elucidated in the context of the bias, where it was shown that in order to derive the bias on large scales one needs to run very small simulations in a curved universe~\cite{Baldauf:2011bh}. This line of reasoning is indeed very similar to what happens in QCD, where we perform lattice simulation to measure quantities relevant at energies above around one GeV, while we use the chiral Lagrangian for predictions at smaller energies.

The effective field theory (EFT) of the long distance universe was initially developed by some of us in~\cite{Baumann:2010tm}. It was noticed that by concentrating on length scales longer than the non-linear scale, the universe is described by a fluid with small perturbations. The equations of motion of this fluid are organized in a derivative expansion in the ratio of the considered wavenumber over the wavenumber associated to the non-linear scale $k_{NL}\sim 1/10$ Mpc$^{-1}$. At leading order in derivatives, the fluid has the stress tensor of an ordinary imperfect fluid, characterized by a speed of sound for the fluctuations, a bulk and a shear viscosity, plus a stochastic pressure component. This makes our approach different with respect to the `standard' approaches both at a quantitative and a qualitative level.

The purpose of this paper is to further develop this effective theory and be able to make observational predictions. The parameters that characterize the fluid, the speed of sounds, bulk viscosity, etc., are determined by the microphysics at the non-linear scale, that we call UV, and cannot be derived from within the effective theory. They have to be either fit to observations, or measured in {\it small} $N$-body simulations. At this point, the EFT becomes predictive. Again, this is very similar to what happens in QCD, where one can measure the pion coupling constant~$F_\pi$ in lattice simulations, after which the Chiral Lagrangian becomes predictive. 

Our basic method and key results are summarized as the following:
\begin{itemize}

\item By smoothing the collisionless Boltzmann equation for non-relativistic matter in an expanding FRW background on a length scale $\Lambda^{-1}$, we establish the continuity and Euler equations for an effective fluid. The Euler equation includes an effective stress-tensor $[\tau^{ij}]_\Lambda$ that is sourced by the short-modes $\delta_s$.

\item By taking correlation functions of the stress tensor in the presence of long wavelength fluctuations, we define an effective stress-tensor that is only a function of the long wavelength fluctuations. It takes the form 
\bea
[\tau^{ij}]_\Lambda\amp=\amp \delta^{ij}p_b+\rho_b\Bigg{[} c_s^2\,\delta^{ij}\delta_l-{c_{bv}^2\over Ha}\delta^{ij}\,\partial_k v_l^k\nonumber\\\amp-\amp{3\over4}{c_{sv}^2\over Ha}\left(\partial^jv_l^i+\partial^iv_l^j-{2\over3}\delta^{ij}\,\partial_kv_l^k\right)
\Bigg{]}+\ldots\ ,
\eea
where the various parameters  $c_s^2,c_{bv}^2$ etc. are defined by proper correlation functions of short wavelength and long wavelength fluctuations.

\item By directly evaluating the stress  tensor  from the microphysical theory, i.e., from $N$-body simulations, and computing the appropriate correlation functions, we calculate the value of the fluid parameters. For a $\Lambda$CDM universe with standard cosmological parameters at redshift $z=0$ and smoothing scale $\Lambda=1/3 h$ Mpc$^{-1}$, we find
\bea
c_{\rm comb}^2{(\Lambda=1/3)} &=0.96 \pm 0.1 \times 10^{-6}\ (c^2)\, ,
\eea
where $c^2_{\rm comb}$ is the combination of $c_s^2, c_{bv}^2$ and $c_{sv}^2$ that is relevant for the leading non-linear correction to the power spectrum (one-loop in perturbation theory), and $c$ is the speed of light.

\item Alternatively, by directly matching the couplings of the effective fluid to the measured power spectrum, we obtain $c_{\rm comb}^2{(\Lambda=1/3)}\simeq  0.9\times 10^{-6}c^2$ in remarkable agreement with the direct measurement from $N$-body simulations. 

\item The fluid parameters carry $\Lambda$ dependence (as does any `bare' parameter in an interacting field theory). This cutoff dependence is taken to cancel against the cutoff dependence of the loop integral. As usual in effective field theories, we `renormalize' the theory by sending the cutoff $\Lambda\to\infty$ and carefully changing the fluid parameters so that predictions at low wavenumbers are not changed in the process.  The finite values of the fluid parameters such as $c_{\rm comb}^2$ in the $\Lambda\to\infty$ limit is a direct measure of the irreducible finite error made in standard approaches that approximate the dark matter on large scales as a pressureless ideal fluid. This is an irreducible error that is not recovered even by solving non-linearly the  equations for a pressureless ideal fluid, as the various perturbative approaches attempt to do. This occurs simply because the equations they solve are not correct. Our approach, in contrast, should reach arbitrary precision, at least in principle, up to non-perturbative corrections.

\item The pressure and viscosity dampen the power spectrum by acting in opposition to gravity, which makes sense intuitively. This is able to help explain the observed shape of the baryon-acoustic-oscillations in the power spectrum relative to standard perturbation theory (SPT).

\item More precisely, at one-loop the density-density power spectrum receives a correction $\delta P$ from the fluid parameters, which we find to be
\beq
\delta P(k)\sim - c_{\rm comb}^2 \frac{k^2}{H^2} P_{11,l}(k)
\label{Pcorr}\eeq
where $P_{11,l}(k)$ is the linear power spectrum. Since this is negative and grows as a function of $k$, the power spectrum is reduced compared to SPT at high $k$'s, improving the  agreement with the full non-linear spectrum.

\item We will find that already at one-loop, the computed power spectrum agrees at percent level with the non-linear one up to $k\sim 0.24 h$ Mpc$^{-1}$. This suggest that in large scale structure surveys we should be able to extract primordial information all the way to at least such an high wavenumber, improving greatly with respect to the CMB our knowledge of the origin of the universe. 

\end{itemize}

During the years there has been a very large and relevant amount of work in understanding perturbatively the large scale clustering of dark matter. An incomplete sample of these works is given by~\cite{Bernardeau:2001qr,Jain:1993jh,Shoji:2009gg,Jeong:2006xd,Crocce:2005xy,Crocce:2007dt,Matsubara:2007wj,McDonald:2006hf,Taruya:2007xy,Izumi:2007su,Matarrese:2007aj,Matarrese:2007wc,Nishimichi:2007xt,Takahashi:2008yk,Carlson:2009it,Fitzpatrick:2009ci,Enqvist:2010ex,Pietroni:2011iz, Tassev:2011ac,Tassev:2012cq,Tassev:2012hu}.

\section{From Dark Matter Particles to Cosmic Fluid}

We take dark matter to be fundamentally described by a set of identical collisionless classical non-relativistic particles interacting only gravitationally. This is a very good approximation for all dark matter candidates apart from very light axions. Given that on large scales baryons follow dark matter, we can include them in the overall dark matter description. As we discuss later, we also neglect general relativistic effects and radiation effects. In this approximation, numerical $N$-body simulations exactly solve our UV theory. The coefficients of our effective fluid can therefore be extracted directly from the $N$-body simulations, following directly the procedure described in~\cite{Baumann:2010tm}. Here, the  UV theory is described by a Boltzmann equation. Therefore, in order to be able to extract the fluid parameters from $N$-body simulations, we need to derive the fluid equations from the Boltzmann equations and subsequently express the parameters of the effective fluid directly in terms of quantities measurable in an $N$-body simulation. This is the task of this section.

\subsection{Boltzmann Equation}

Let us start from a one-particle phase space density $f_n(\vec x,\vec p)$ such that $f_n(\vec x,\vec p) d^3x d^3p$ represents the probability for the particle $n$ to occupy the infinitesimal phase space volume $d^3xd^3p$. For a point particle, we have
\be
f_{n}(\vec x,\vec p)=\delta^{(3)}(\vec x-\vec x_n)\delta^{(3)}(\vec p-m\, a\, \vec v_n)\ .
\ee
The total phase space density $f$ is defined such that $f(\vec x,\vec p) d^3xd^3p$ is the probability that there is a particle in the infinitesimal phase space volume $d^3xd^3p$:
\be
f(\vec x,\vec p)=\sum_n\delta^{(3)}(\vec x-\vec x_n)\delta^{(3)}(\vec p-m\,a\,\vec v_n)\ .
\ee
We define the mass density $\rho$, the momentum density $\pi^i$ and the kinetic tensor $\sigma^{ij}$ as
\bea
&&\rho(\vec x,t)=\frac{m}{a^3}\int d^3p\; f(\vec x,\vec p)= \frac{m}{a^3}\sum_n \delta^{(3)}(\vec x-\vec x_n)\ , \\
&&\pi^i(\vec x,t)=\frac{1}{a^4}\int d^3p\; p^i  f(\vec x,\vec p)= \frac{m}{a^3}\sum_n v_n^i \delta^{(3)}(\vec x-\vec x_n)\ ,\\ \nonumber
&&\sigma^{ij}(\vec x,t)=\frac{1}{m a^5}\int d^3p\; p^i p^j  f(\vec x,\vec p)=\sum_n \frac{m}{a^3} v^i_n v^j_n\delta^{(3)}(\vec x-\vec x_n)\ .
\eea
The particle distribution $f_n$ evolves accordingly to the Boltzmann equation
\be
\frac{D f_n}{D t}=\frac{\d f_n}{\d t}+\frac{\vec p}{m a^2}\cdot \frac{\d f_n}{\d \vec x}-m \sum_{\bar n\neq n}\frac{\d \phi_{\bar n}}{\d\vec x}\cdot\frac{\d f_n}{\d\vec p}=0\ ,
\ee
where $\phi_n$ is the single-particle Newtonian potential.
There are two important points to highlight about the former equation. First, we have taken the Newtonian limit of the full general relativistic Boltzmann equation. This is an approximation we make for simplicity. All our results can be trivially extended to include general relativistic effects. However, it is easy to realize that the Newtonian approximation is particularly well justified. Non-linear corrections to the evolution of the dark matter evolution are concentrated at short scales, with corrections that scale as $k^2/k_{NL}^2$. General relativistic corrections are expected to scale as $k^2/(aH)^2$. This means that we should be able to cover up to wavelength of order $300$ Mpc before worrying about per mille General relativity corrections. Furthermore, one of the main goals of this paper is to recover the parameters of the effective fluid of the universe from very short scale simulations valid on distances of order of the non-linear scale. The parameters we will extract in the Newtonian approximation are automatically valid also for the description of an effective fluid coupled to gravity in the full general relativistic setting.

A second important point to highlight in the former Boltzmann equation is about the single-particle Newtonian potential $\phi_n$. Following~\cite{Baumann:2010tm}, the Newtonian potential $\phi$ is defined through the Poisson equation
\be
\d^2\phi=4\pi G a^2 \left(\rho-\rho_b\right)\ ,
\ee
with $\rho_b$ being the background density and $\d^2=\delta^{ij}\d_i\d_j$. We raise and lower spatial indexes with $\delta_{ij}$.  The solution reads
\bea
&&\phi=\sum_n \phi_n+\frac{4\pi G a^2\rho_b}{\mu^2}\ ,\\
&&\phi_n(\vec x)=-\frac{ G\,m}{|\vec x-\vec x_n|} e^{-\mu|\vec x-\vec x_n|}\ .
\eea
Notice that the overall $\phi(\vec x)$ is IR divergent in an infinite universe. This is due to a breaking of the Newtonian approximation. We have regulated it with an IR cutoff $\mu$ that we will take to zero at the end of the calculation. Our results do not depend on $\mu$, as indeed we are interested in very short distance physics.

By summing over $n$, we obtain the Boltzmann equation for $f$
\be
\frac{D f}{D t}=\frac{\d f}{\d t}+\frac{\vec p}{m a^2}\cdot \frac{\d f}{\d \vec x}-m \sum_{n,\bar n;\bar n\neq n}\frac{\d \phi_{\bar n}}{\d\vec x}\cdot\frac{\d f_n}{\d\vec p}=0\ .
\ee

\subsection{Smoothing}

Following~\cite{Baumann:2010tm}, we construct the equations of motion for the effective fluid by smoothing the Boltzmann equations and by taking moments of the resulting long-distance Boltzmann equation. The smoothing guarantees that the Boltzmann hierarchy can be truncated, leaving us with an effective fluid. indeed, notice that it is not trivial at all that we should end up with an effective fluid. Fluid equations are usually valid over distances longer than the mean free path of the particles. But here for dark matter particles the mean free path is virtually infinite. What saves us is that the dark matter particles have had a finite amount of proper time, of order $H^{-1}$,  to travel since reheating, and they traveled at a very non-relativistic speed. This defines a length scale $v H^{-1}\sim 1/k_{NL}$ which is indeed of order of the non-linear scale. This length scale plays the role of a mean free path, as verified in~\cite{Baumann:2010tm}. The truncation of the Boltzmann hierarchy is regulated by powers $k/k_{NL}\ll 1$.
 
We define the Gaussian smoothing
\be
W_\Lambda(\vec x)=\left(\frac{\Lambda}{\sqrt{2\pi}}\right)^3 e^{-\frac{1}{2}\Lambda^2 x^2}\ , \qquad W_\Lambda(k)=e^{-\frac{1}{2}\frac{k^2}{\Lambda^2}}\ ,
\ee
with $\Lambda^2$ representing a $k$-space, comoving cutoff scale. This will smooth out quantities with wavenumber $k\gtrsim \Lambda$, or equivalently with waveleghts smaller than $\lambda\lesssim1/\Lambda$. We regularize our observable quantities ${\cal O}(\vec x,t)$, $\rho,\pi,\phi,\ldots$ , by taking convolutions in real space with the filter, defining long-wavelength quantities as
\be
{\cal O}_l(\vec x,t)=\left[{\cal O}\right]_\Lambda(\vec x,t)=\int d^3x' \,W_\Lambda(\vec x-\vec x') {\cal O}(\vec x')\ .
\ee
Notice that in Fourier space $W(k)\rightarrow 1$ as $k\rightarrow 0$: our fields are asymptotically untouched at long distances. 

The smoothed Boltzmann equation becomes
\be
\left[\frac{D f }{D t}\right]_\Lambda=\frac{\d f_l}{\d t}+\frac{\vec p}{m a^2}\cdot\frac{\d f_l}{\d \vec x}-m\sum_{n,\bar n,n\neq \bar n}\int d^3x' W_\Lambda(\vec x-\vec x')\frac{\d\phi_n}{\d\vec x'}(\vec x')\cdot\frac{\d f_{\bar n}}{\d\vec p}\ .
\ee
Fluid equations are obtained by taking successive moments
\be
\int d^3p\; p^{i_{1}}\ldots p^{i_n} \left[\frac{D f}{D t}\right]_\Lambda(\vec x,\vec p)=0\ ,
\ee 
creating in this way a set of coupled differential equations known as Boltzmann hierarchy. As we will explain in more detail later, it will be sufficient for the purposes this paper to stop at the first two moments (one-loop approximation). The first  two moments will give the continuity and momentum equations in the approximation in which the fluid is described by the Navier-Stokes approximation, with the addition of a stochastic term. We obtain
\bea\label{eq:continuity}
&&\dot\rho_l+3 H\rho_l+\frac{1}{a}\d_i(\rho_l v_l^i)=0\  ,\\ \label{eq:momentum}
&&\dot v_l^i+ H v_l^i+\frac{1}{a}v_l^j\d_j v_l^i+\frac{1}{a}\d_i\phi_l=-\frac{1}{a \rho_l}\d_j\left[\tau^{ij}\right]_\Lambda\ .
\eea
Let us define the various quantities that enter in these equations. We define the long wavelength velocity field as the ratio of the momentum and the density
\be
v_l^i=\frac{\pi_l^i}{\rho_l}\ .
\ee
The right hand side of the momentum equation (\ref{eq:momentum}) contains the divergence of an effective stress tensor which is induced by the short wavelength fluctuations. This is given by
\be
\left[\tau^{ij}\right]_\Lambda=\kappa^{ij}_l+\Phi_l^{ij}\ ,
\ee
where $\kappa$ and $\Phi$ correspond to `kinetically-induced' and `gravitationally-induced' parts:
\bea\label{eq:short_stress}
&&\kappa^{ij}_l=\sigma_l^{ij}-\rho_l v_l^i v_l^j\ ,\\ \nonumber
&&\Phi_l^{ij}=-\frac{1}{8\pi G a^2}\left[w_l^{kk}\delta^{ij}-2w_l^{ij}-\d_k\phi_l\d^k\phi_l\delta^{ij}+2\d^i\phi_l\d^j\phi_l\right]\ ,
\eea
where
\be
w_l^{ij}(\vec x)=\int d^3x' W_\Lambda(\vec x-\vec x')\left[\d^i \phi(\vec x')\d^j\phi(\vec x')-\sum_n \d^i\phi_n(\vec x')\d^j\phi_n(\vec x')\right]\ .
\ee
Note that we have subtracted out the self term from $w_l^{ij}$, as necessary when passing from the continuous to the discrete description in the Newtonian approximation, and used that $\d^2\phi=4\pi G a^2(\rho-\rho_b)$ and  $\d^2\phi_l=4\pi G a^2(\rho_l-\rho_b)$ to express $\Phi_l$ in terms of $\phi$ and $\phi_l$.  In the limit in which there are no short wavelength fluctuations, and $\Lambda\rightarrow \infty$, $\kappa_l$ and $\Phi_l$ vanish. In App.~\ref{app:short_modes} we provide the above expression written just in terms of the short wavelength fluctuations.

\subsection{Integrating out UV Physics}

The effective stress tensor that we have identified is explicitly dependent on the short wavelength fluctuations. These are very large, strongly coupled, and  therefore impossible to treat within the effective theory. When we compute correlation functions of long wavelength fluctuations, we are taking expectation values. Since short wavelength fluctuations are not observed directly, we can take the expectation value over their values. This is the classical field theory analog of the operation of `integrating out' the UV degrees of freedom in quantum field theory, now applied to classical field theory. The long wavelength perturbations will affect the result of the expectation value of the short modes, through, e.g., tidal like effects. This means that the expectation value will depend on the long modes. In practice, we take the expectation value on a long wavelength background. The resulting function depends only on long wavelength fluctuations as degrees of freedom. In this way, we have defined an effective theory that contains only long wavelength fluctuations.  Since long wavelength fluctuations are perturbatively small,  we can Taylor expand in the size of the long wavelength fluctuations. Schematically we have
\be
\langle\left[\tau^{ij}\right]_\Lambda\rangle_{\delta_l}=\langle\left[\tau^{ij}\right]_\Lambda\rangle_{0}+\left.\frac{\d\langle\left[\tau^{ij}\right]_\Lambda\rangle_{\delta_l}}{\d\delta_l}\right|_0 \delta_l+\ldots\ .
\ee
For the precision we pursue in the rest of the paper, we will stop at linear level in the long wavelength fluctuations, though nothing stops us from going to higher order. By the symmetries of the problem, the resulting stress tensor must take the following form
\be\label{eq:effectve_stress}
\langle\left[\tau^{ij}\right]_\Lambda\rangle_{\delta_l}=p_b\delta^{ij}+\rho_b \left[c_s^2\delta_l \delta^{ij}-\frac{c_{bv}^2}{H a}\delta^{ij}\d_k v_l^k-\frac{3}{4}\frac{c_{sv}^2}{H a}\left(\d^j v_l^i+\d^i v_l^j-\frac{2}{3}\delta^{ij}\d_k v^k_l\right)\right]+\Delta\tau^{ij}+\ldots\ .
\ee
This is the stress tensor of an imperfect fluid. $p_b$ is the background pressure that is induced by short distance inhomogeneities even in the absence of long wavelength fluctuations. $c_{s}^2$ is the speed of sounds of the fluctuations: $\delta p=c_s^2\delta\rho$. The parameters $c_{bv}$ and $c_{sv}$ are the coefficients for the  bulk $\zeta$ and the shear $\eta$ viscosity respectively, with units of velocity. They are related to $\eta$ and $\zeta$ by the relation
$\eta=3\rho_b c_{sv}^2/(4H),\  \zeta=\rho_b c_{bv}^2/H$ .
$\Delta\tau^{ij}$ represents a stochastic term, that takes into account the difference between the actual value of $\tau^{ij}$ in a given realization and its expectation value~\footnote{For the readers familiar with the in-in formalism, this term will take into account the cut-in-the-middle one-loop diagrams~\cite{Senatore:2009cf}.}. We will come back to this term shortly, but it is worth noting that neglecting this term in the above equations reproduces the familiar Navier-Stokes equations.

Finally, the ellipses ($\ldots$) represent terms that are either higher order in $\delta_l$, or higher order on derivatives of $\delta_l$. Indeed, higher derivative terms will be in general suppressed by $k/k_{NL}\ll 1$, and, as typical in effective field theories, we take a derivative expansion in those. Astrophysically, these terms would corresponds to the effects induced by a sort of higher-derivative tidal tensor. Once we expand in derivatives of the long wavelength fluctuations, we take the parameters in (\ref{eq:effectve_stress}) to be spatially independent, but time dependent.

The coefficient $\delta p_b,c_s,c_{sb},c_{sv}$ are determined by the UV physics and by our smoothing cutoff $\Lambda$, and are not predictable within the effective theory. They must be measured from either $N$-body simulations, or fit directly to observations. This is akin to what happens in the Chiral Lagrangian for parameters that can be measured in experiments or in lattice simulations, such as $F_\pi$. We first define the correlation functions that will allow us to extract these parameters from small $N$-body simulations.

\subsection{Matching Correlation Functions}

It useful to define the following quantities from the stress tensor
\bea
 J^i_l=\frac{1}{a\rho_b}\d_j\left[\tau^{ij}\right]_\Lambda\ ,&& \qquad\qquad A_l^{ki}=\frac{1}{a}\d^k J_l^i\ , \\ \nonumber
 A_l=\frac{1}{a}\d_i J^i_l\ ,&& \qquad\qquad B_l=\frac{1}{a^2\rho_b}\left(\d_i\d_j-\delta_{ij}\d^2\right)\left[\tau^{ij}\right]_\Lambda\ ,
\eea
and to introduce a dimensionless velocity divergence
\be
\Theta_l=-\frac{\d_k v_l^k}{H a}\ ,\qquad \Theta_l^{ki}=-\frac{\d^k v_l^i}{H a}\ .
\ee
Then, according to (\ref{eq:effectve_stress}), we have
\bea
&& a J^i_l=c_s^2\d_i\delta_l+\frac{3}{4}c_{sv}^2\d_j\Theta_l^{ji}+\left(\frac{c_{sv}^2}{4}+c_{bv}^2\right)\d_i\Theta_l\ ,\\ \nonumber
&& a^2 A_l^{ki}=c_s^2\d^k\d^i\delta_l+\frac{3}{4}c_{sv}^2\d^k\d_j\Theta_l^{ji}+\left(\frac{c_{sv}^2}{4}+c_{bv}^2\right)\d^k\d^i\Theta_l\ ,\\ \nonumber
&& a^2 A_l=c_s^2\d^2\delta_l+\left(c_{sv}^2+c^2_{bv}\right)\d^2\Theta_l\ ,\\ \nonumber
&& a^2 B_l=c_{sv}^2 \d^2\Theta_l\ .
\eea
In order to extract the parameters of the effective fluid, we multiply each of these functions with long wavelength fields, and take expectation values. By forming suitable combinations of these, the parameters of the effective fluid can be extracted. We will need the following set of correlation functions
\bea\label{eq:correlations}
&& P_{A\delta}(x)=\langle A_l (\vec x'+\vec x)\delta_l(\vec x')\rangle\ ,\\  \nonumber
&& P_{A\Theta}(x)=\langle A_l (\vec x'+\vec x)\Theta_l(\vec x')\rangle\ ,\\  \nonumber
&& P_{A^{ki}\Theta_{ki}}(x)=\langle A_l^{ki} (\vec x'+\vec x)\Theta_l{}_{ki}(\vec x')\rangle\ ,\\  \nonumber
&& P_{B\Theta}(x)=\langle B_l (\vec x'+\vec x)\Theta_l(\vec x')\rangle\ ,\\  \nonumber
&& P_{\delta\delta}(x)=\langle \delta_l (\vec x'+\vec x)\delta_l(\vec x')\rangle\ ,\\  \nonumber
&& P_{\delta\Theta}(x)=\langle \delta_l (\vec x'+\vec x)\Theta_l(\vec x')\rangle\ ,\\  \nonumber
&& P_{\Theta\Theta}(x)=\langle \Theta_l (\vec x'+\vec x)\Theta_l(\vec x')\rangle\ ,\\  \nonumber
&& P_{\Theta^{ji}\Theta^{k}_i}(x)=\langle \Theta_l^{ji} (\vec x'+\vec x)\Theta_l^{k}{}_i(\vec x')\rangle\ ,\\  \nonumber
\eea
where $\delta_l=\delta\rho_l/\rho_b$.
From these we obtain the following expressions for the parameters of the effective theory
\bea\label{eq:parameters}
&& c_s^2=a^2 \frac{P_{A\Theta}(x)\d^2P_{\delta\Theta}(x)-P_{A\delta}(x)\d^2P_{\Theta\Theta}(x)}{\left(\d^2P_{\delta\Theta}(x)\right)^2-\d^2P_{\delta\delta}(x)\d^2P_{\Theta\Theta}(x)}\ , \\ \nonumber
&& c_{v}^2=a^2 \frac{P_{A\delta}(x)\d^2P_{\delta\Theta}(x)-P_{A\Theta}(x)\d^2P_{\delta\delta}(x)}{\left(\d^2P_{\delta\Theta}(x)\right)^2-\d^2P_{\delta\delta}(x)\d^2P_{\Theta\Theta}(x)}\ , \\ \nonumber
&& c_{sv}^2=\frac{4}{3} a^2 \frac{P_{A^{ki}\Theta_{ki}}(x)-P_{A\Theta}(x)}{\d^2P_{\Theta^{ki}\Theta_{ki}}(x)-\d^2P_{\Theta\Theta}(x)}=a^2 \frac{P_{B\Theta}(x)}{\d^2 P_{\Theta\Theta}(x)}\ , \\ \nonumber
\eea
where $c_v^2=c_{sv}^2+c_{bv}^2$ is the sum of the viscosity coefficients. By extracting the correlation functions in (\ref{eq:correlations}) from $N$-body simulations, and performing the ratios in (\ref{eq:parameters}), we should be able to extract the parameters of the effective theory. Notice that the ratios are supposed to be spatially independent. Such behavior is expected to hold at large distances $x\gg \Lambda^{-1}$ where the higher derivative terms are negligible.

In simulations, we should in principle also measure the stochastic components of the stress tensor. In the two point function at one-loop, at leading order in derivatives, it enters just the correlation function of the trace. This amounts to measuring
\be
\langle J^i_l(\vec x,t)J^j_l(\vec x+\vec x',t')\rangle\ .
\ee
We will see that the effect of this stochastic term is accidentally higher order in $\delta_l$ and so does not enter at leading order. 

After all these parameters have been measured in $N$-body simulations, the EFT is prone for perturbation theory. It is alternatively possible to perform directly perturbation theory and fit the results to observations. We will be able to perform both approach and check that we obtain the same result. We will describe in detail how to measure these quantities in simulations in App.~\ref{app:simulations} and we will give the results of these measurements in Sec.~\ref{sec:simulation_results}.  For the moment we will instead directly move to apply perturbation theory with our~EFT.

These parameters can either be measured from $N$-body simulations directly or kept generic and then extracted by fitting the results to  observables. We are able to do both and verify that we obtain the same result. We will describe in detail how to measure these quantities in simulations in App.~\ref{app:simulations} and we will give the results of these measurements in Sec.~\ref{sec:simulation_results}.  First we develop perturbation theory within the EFT that will allow us to make predictions and to extract  these parameters from observations.

\section{Perturbation Theory with the EFT}

We now proceed to perform perturbation theory within our EFT. The non-linear equations of motion that we need to solve are
\bea\label{eq:all_equations}
&&\nabla^2\phi_l=\frac{3}{2}{H}_0^2\Omega_m \frac{a_0^3}{a}\delta_l+\dots \ ,\\ \nonumber
&&\dot\delta_l=-\frac{1}{a}\d_i\left((1+\delta_l) v_l^i\right)\ ,\\ \nonumber
&&\dot{v}_l^i+H v_l^i+\frac{1}{a} v_l^j\d_jv_l^i+\frac{1}{a}\d^i\phi_l=-\frac{1}{a}c_s^2\d^i\delta_l+\frac{3}{4}\frac{c_{sv}^2}{H a^2}\d^2 v_l^i+\frac{4 c_{bv}^2+c_{sv}^2}{4 H a^2}\d^i\d_j v_l^j-\Delta J^i+\ldots\ ,
\eea
where $\dot{}=d/dt$, $H=\dot{a}/a$, $\Delta J^i=\d_j(\Delta \tau^{ji})/(a\rho_b)$, $\Omega_m$ is the present day matter fraction, $a_0$ is the present day scale factor, usually taken to be equal to 1, and $\ldots$ represent higher order terms (in  sense that we will explain shortly) that come from the expression of the short wavelength stress tensor $\tau^{ij}$ in terms of long wavelength fluctuations. Our theory is defined on scales longer than the non-linear scale. For this reason we have $\delta_l\ll1$. The long-wavelength velocity $v_l$ and the long-wavelength $\phi_l$ are small even inside the the non-linear scale and they are even smaller at larger distances. This means that we can reliably solve the above non-linear equations iteratively around the linear solution. Such an iterative solution is very similar to what is done in quantum field theory, where the solution to the quantum non-linear equations is organized in Feynman diagrams. Indeed we can organize the various perturbative terms around Feynman diagrams even in this case. The result is very similar to what is computed in the in-in formalism, for example when computing quantum corrections to inflationary correlation functions~\cite{Senatore:2009cf}. Indeed, the calculation we are going to do shares many of the features that are present in normal quantum field theory computations: cutoff, renormalization, running, and so on are all concepts that will appear and prove useful as we proceed. They have nothing to do with the word `quantum' in `quantum field theory', rather they have to do with the `field theory'. Our calculation is for a classical field theory and shares all these features.

\subsection{Organization of the perturbation theory}

The simplest way to organize our perturbation theory is to use the fact that, in any order of magnitude approximation, $\phi$ is constant at all scales, of order $10^{-5}$, and that well inside the horizon the Newtonian approximation holds. For length scales longer than the equality scale, at the linear level we therefore have
\bea\label{eq:scaling}
&&\phi_l\left(\Delta x\sim L\right)\sim10^{-5}\ , \qquad v_l\left(\Delta x\sim L\right)\sim  10^{-5}\frac{1}{H L} \ , \\ \nonumber
&&\delta_l\left(\Delta x\sim L\right)\sim \frac{1}{H^2\d^2}\phi_l\left(\Delta x\sim L\right)\sim 10^{-5}\frac{1}{H^2L ^2}\ . 
\eea
We see that as $L\rightarrow 0$, $\delta_l$ grows and indeed becomes of order one at $L\sim\lambda_{NL}$~\footnote{For the propose of estimating at order of magnitude level, we have assumed that the $k$ modes under consideration are longer than the equality scale $k_{\rm eq}\sim 0.01$ Mpc$^{-1}$ and taken $\phi_l$ to be $k$-independent.  On shorter scales $\phi_l$ decays, and the estimates need to be slightly modified. This subtlety will be important for the actual numerical contribution of the various terms, and it will be properly accounted for, but is not particularly relevant for the order of magnitude estimates that control our power counting, and so we will ignore it for simplicity's sake.}. At distances larger than the non-linear scale, we therefore expand in powers of $\delta_l$, keeping in mind that the additional fluctuations scale as in (\ref{eq:scaling}).
Let us estimate the relative size of the terms.

{\bf Loop corrections:} it is easy to estimate from the non linear structure of the equations that $\d_i v^i\sim H\delta_l$. Notice that in the power spectrum we need to take two non-linear corrections, or alternatively look at the cubic corrections. The non-linear terms scale as
\be
\frac{{\rm non}-{\rm linear\ terms}}{\rm Hubble\ friction}\sim\frac{\delta_l v^j_l\d_j v^i_l}{H v^i_l}\sim \frac{H\ \delta_l^2 v^i_l}{H v^i_l}\sim \delta_l^2
\ee
Loop corrections therefore scale as $\delta_l^2$, peaked at the highest possible scale within the theory~$\Lambda$.

{\bf Pressure and Viscosity terms:} these terms result from integrating out the modes higher than the $\Lambda$ scale. So, naively they should scale as the $\delta$ at the extreme UV scales beyond the effective theory. Since the theory in the UV is strongly coupled, very large corrections are expected, and the result cannot be extrapolated from the linear regime, even at the order of magnitude level. This is why we will measure the parameters such as $c_s^2$ from $N$-body simulations. What we will find is that these parameters are of order $10^{-5}$. This happens because the combination of short modes that generates the parameters like $c_s$ are such that short modes that have virialized do not contribute~\footnote{This was found and used in~\cite{Baumann:2010tm} to show that there is very little backreaction on the evolution of the universe from short scale non-linearities. Short scale gravitational collapse changes the equation of state of the overall universe by a relative factor of order $10^{-5}$.}. Since these terms scale as $\phi\delta$, the contribution is peaked at those modes that have just become non-linear $\delta\sim1$, but not yet virialized. We therefore expect the parameters $c_s^2$'s to be of order $\phi\sim 10^{-5}$. We will see that the fact that short modes entered the horizon in the radiation era makes this a bit of an overestimate. At this point we are ready to estimate  the size of these corrections:
\be
\frac{\rm Pressure\,,Viscosity}{\rm Hubble\ friction}\sim\frac{c_s^2 \d \delta_l}{H v^i_l}\sim c_s^2\frac{ \d^2\delta_l}{H^2 \delta_l}\sim \frac{c_s^2}{10^{-5}}\delta_l\ .
\ee
Notice that thanks to the strongly coupled UV theory (or thanks to the fact that non-linear structures virialize), we have that for $c_s^2\sim 10^{-5}$, the contribution from these terms is larger than the one loop contribution in the low energy theory. This is so because the theory is strongly coupled in the UV. We conclude that one insertion of these terms counts at least as a one-loop term.

{\bf Stochastic terms:} Let us now continue on to evaluate the effect of the stochastic terms. This is a bit more complicated. Let us evaluate the relative effect on the power spectrum. The structure of the equations leads to the following approximate non-linear solution
\be
\delta_{l,non-lin.}\sim\delta_{l,lin}+c_s^2\frac{\d^2}{H^2}\delta_{l,lin}+\frac{\d^2}{H^2}\frac{\Delta \tau}{\rho_b}\ .
\ee
In the power spectrum we therefore have
\be
\langle\delta_l\delta_l\rangle_{\rm 1-loop}\sim c_s^2\frac{k^2}{H^2}\langle\delta^2_l\rangle+ \left(\frac{ k^2}{H^2\rho_b}\right)^2\langle\Delta\tau^2\rangle\ ,
\ee
as the stochastic part must be correlated with itself. Due to virialization, we expect that the correlation function of $\tau$ should be Poisson like on independent pixels of order the non-linear scale $k_{NL}^{-1}$. We therefore estimate
\be
\langle\Delta\tau^2\rangle\sim\langle\tau^2\rangle \left(\frac{k}{k_{NL}}\right)^3\sim \left(c_s^2\rho_b \right)^2 \left(\frac{k}{k_{NL}}\right)^3\ .
\ee
This is indeed confirmed by calculations in perturbation theory~\cite{Baumann:2010tm}. We therefore have
\be
\frac{\rm Stochastic}{\rm Pressure}\sim\frac{k}{k_{NL}}\ ,\quad \Rightarrow\quad \frac{\rm Stochastic}{\rm Friction}\sim\frac{k}{k_{NL}}\delta_l\sim \delta_l^{3/2}\ .
\ee
This tells us that on scales longer than the non-linear scale, the contribution of the stochastic pressure is parametrically smaller than the pressure effects. Since in this paper we will stop at one-loop order, we can therefore neglect this correction. It should be noted that these contributions scale parametrically differently than the loop contributions, and so, depending on the scale considered, they might be more relevant that a 2-loop contribution. We notice that something similar happens also at the level of dissipative fluids, where we generically include dissipative terms through the Navier-Stokes equations, but we neglect stochastic terms.

{\bf Higher Derivative Terms:} When we take the expectation value of the short-distance stress tensor in a background of a long mode, we have Taylor expanded both in the long wavelength fluctuations and in their derivatives. Higher power corrections scale with powers of $\delta_l$, and so corresponds to higher-loop terms. Higher derivative terms instead scale as powers of $k^2/k_{NL}^2\sim\delta_l$, where we have taken the squared because of rotational invariance. This shows that higher derivative terms scale nicely as loop terms. If we allow the cutoff to remain finite, we should also include higher derivative terms that scale as $k^2/\Lambda^2$.

{\bf General Relativistic and Radiation Corrections:} In this paper we will neglect general relativistic corrections and all non-linear contributions coming from the fact that the universe was radiation dominated at early times. General Relativistic corrections scale as
\be
\frac{\rm GR~Corrections}{\rm Newtonian~Approximation}\sim\left(\frac{H}{k}\right)^2\sim \frac{10^{-5}}{\delta_l}
\ee
For the high scales where non-linear corrections are relevant, for example at the BAO scale $k_{BAO}\sim~10^{-2}$, these corrections are of order $10^{-4}$, and so uninteresting from this point of view. 
Radiation is the dominant component of the universe at early times. Neglecting it amounts to neglect corrections that scale as $a/a_{eq}\sim 10^{-3}$, where $a$ is the scale factor and $a_{eq}$ is the scale factor at matter radiation equality. Inclusion of these corrections in perturbation theory has been studied in~\cite{Fitzpatrick:2009ci}, and it gives a small correction to power spectrum, and small, but potentially measurable, corrections to the three-point function, corresponding to $f_{NL}\sim$few.
Both General Relativistic effects and radiation effects do not represent an intrinsic limitation of our EFT. They can be straightforwardly included in our formalism, by simply improving the equations of motion we use in this paper. 

In summary, we see that apart from the stochastic terms and some higher derivative terms, all the remaining terms: loops, pressure, higher derivatives and higher powers of $\delta_l$ from $\tau$,  scale as powers of $\delta_l$, which is our main ordering parameter. Stochastic terms instead contribute at leading order  as $\delta_l^{3/2}$, so they count as one loop and a half. Cutoff-dependent higher derivative terms scale as $k^2/\Lambda^2$.

{\bf Cutoff dependence and effective expansion parameter:} So far, it looks like that our expansion parameter is the highest $\delta_l$ we have in our theory, which is $\delta_l(k\sim \Lambda)\sim \Lambda^2/k_{NL}^2$. However, the situation is even better than this. So far, we have defined our theory with a regulating cutoff at $k\sim \Lambda$. Because of this, all our intermediate results depend explicitly on~$\Lambda$:~$c_s(\Lambda),\, c_{sv}(\Lambda),$ etc. and loops need to be cutoff at $\Lambda$. This induces an explicit $\Lambda$ dependence plus higher derivative terms of order $k^2/\Lambda^2$. However, the sum of all the diagrams will be independent of $\Lambda$. Indeed $c_s(\Lambda),$ etc. should be really thought of as one-loop counterterms.  Upon carefully choosing the counterterms $c_s(\Lambda)$, etc., any $\Lambda$ dependence cancels apart from terms in $k/\Lambda$ that should be removed by higher derivative corrections  in the stress tensor that we neglected. In order to resolve such error with the least effort, we will choose the counterterms at a fixed cutoff in such a way as to have the theory agree with observations at a certain renormalization scale $k_{\rm ren.}$, and then we will extrapolate our results to $\Lambda\rightarrow\infty$, effectively letting the residual terms  in $k/\Lambda$ vanish. In this $\Lambda\to\infty$ regime, the loop term gets dominated by the regime in which one of the modes has wavenumber of order the non-linear scale, while the other has a wavenumber of order of the external wavenumber. In this way, loop terms scale as one-power of $\delta_l$ as the counterterm~\footnote{Another kinematically allowed possibility  is for the modes to have both  wave numbers close to the non-linear scale, but slightly different so that their sum is equal to the external wavenumber $\vk$. This contribution would naively scale as $(\delta_l)^0\sim 1$. However this contribution in this regime scales as the stochastic term $\delta_l^{3/2}$.}. At this point, the expansion parameter of the EFT will be $\delta_l\sim k^2/k_{NL}^2$ evaluated at the scale of the external modes, with no residual $\Lambda$ dependence even in the expansion parameters.

Again, this is very similar to what happens when one computes loop corrections in the Chiral Lagrangian. After regulating the chiral theory with a cutoff $\Lambda$, there are naively two expansion parameters. If $E$ is the energy scale of the process, we have $E/F_\pi$ and $E/\Lambda$. After renormalization and by sending $\Lambda\rightarrow \infty$, we are left only we $E/F_\pi$ as an expansion parameters.

\vspace{0.4cm}
 In summary, the expansion parameter of the EFT is $\delta_l^{1/2}\sim (k/k_{NL})$, where $k$ is the typical wavenumber of the external modes. Loops in the EFT, counterterms and higher-derivative terms scale as $\delta_l$. Stochastic terms start contributing at order $\delta_l^{3/2}$.

\subsection{One-loop Perturbation Theory}

We are now ready to implement perturbation theory for the power spectrum at quartic order in $\delta_l$, that is at one-loop. At this order, the equations we are going to solve are the ones in~(\ref{eq:all_equations}) with $\Delta J$ and the $\ldots$ terms neglected.

Let us write the equation for the vorticity $w_l^i=\epsilon^{ijk}\d_j v_k$. Neglecting the stochastic terms that we argued are small, we have
\be
\left(\frac{\d}{\d t}+H-\frac{3 c_{sv}^2}{4 H a^2}\d^2\right)w^i_l=\epsilon^{ijk}\d_j\left(\frac{1}{a}\epsilon_{kmn}v^m_l w_l^n\right)\ .
\ee
In linear perturbation theory the vorticity is driven to zero, and this occurs even the more so at this order in perturbation theory, as the source is proportional to $w_l$. While at higher order one could expect vorticity to be generated, at this order, and therefore for the purposes of this paper, we can take it to be zero. This means that we can work directly with the divergence of the velocity
\be
\theta_l=\d_i v^i_l
\ee
Using $a$ as our time variable, the equations (\ref{eq:all_equations}) reduce to
\bea\label{eq:master}
&&a{\cal H} \delta_l'+\theta_l=-\int \frac{d^3q}{(2\pi)^3}\alpha(\vec q,\vec k-\vec q)\delta_l(\vec k-\vec q)\theta_l(\vec q)\ , \\ \nonumber
&&a {\cal H} \theta_l'+{\cal H} \theta_l+\frac{3}{2}\frac{{\cal H}_0^2\Omega_m}{a} \delta_l-c_s^2 k^2 \delta_l+\frac{c_v^2 k^2}{{\cal H}}\theta_l=-\int \frac{d^3q}{(2\pi)^3}\beta(\vec q,\vkkp)\theta_l(\vec k-\vec q)\theta_l(\vec q)\ ,
\eea
where ${\cal H}=a^{-1}\d a/\d\tau$, subscript $_0$ for a quantity means that the quantity is evaluated at present time, we have set $a_0=1$, $'$ represents $\d/\d a$ and
\be
\alpha(\vec k,\vec q)=\frac{\left(\vec k+\vec q\right)\cdot\vec k}{k^2}\ ,\qquad\beta(\vec k,\vec q)=\frac{\left(\vec k+\vec q\right)^2 \vec k\cdot\vec q}{2 q^2\vec k^2}\ .
\ee
As we discussed, the parameters $c_s,\,c_{bv}$ and $c_{sv}$ are time dependent and must be measured in the simulations as a function of time.  For the purposes of this paper, we will make the simplifying assumption that their time dependence can be inferred in perturbation theory. In other words, we will measure them at one time and deduce their values at different times by perturbation theory~\footnote{As we will see, these parameters need to cancel the $\Lambda$ dependence associated to the regularized loops. The part of these parameters that depends on $\Lambda$ can be therefore reliably inferred in perturbation theory. However, the part of these parameters that is $\Lambda$-independent and that represents the finite contributions should be measured in simulation or in observation. We will assume that the time dependence for these two components is the same. We will check that this is an accurate approximation in an upcoming paper~\cite{foreman}. For this approximation, we stress that since these are 1-loop terms, it is important to know them up to a relative factor of order $\delta_l\ll1$.}. 

\subsubsection{Perturbative Solutions}

Since the correlation function of matter overdensities is small at large distances, we can solve the above set of equations (\ref{eq:master}) perturbatively in the amplitude of the fluctuations. For the computation of the power spectrum at one loop, it is enough to solve these equations iteratively up to cubic order. Order by order, the solution is given by convolving the  retarded Green's function associated to the linear differential operator with the non-linear source term evaluated on lower order solutions. At second order we obtain
\bea\label{eq:delta2}
&&\delta^{(2)}_l(\vec k,a)=\frac{1}{16\pi^3 D(a_0)^2}\\ \nonumber
&&\left[\left(\int_0^a d\tilde a\,G(a,\tilde a)  \tilde a^2  {\cal H}^2(\tilde a) D'(\tilde a)^2\right) \left(2 \int d^3q \beta(\vec q,\vec k-\vec q) \delta s_1(\vec k-\vec q) \delta s_1(\vec q)\right)\right.\\ \nonumber 
&&\left.+\left(\int_0^a d\tilde a\,G(a,\tilde a)\left(2 \tilde a^2 {\cal H}^2(\tilde a) D'(\tilde a)^2 + 3 {\cal H}_0^2\Omega_m \frac{D(\tilde a)^2}{\tilde a} \right)\right) \right.\\ \nonumber
&&\qquad \qquad\qquad\qquad\qquad\qquad \times\left. \left( \int d^3q \alpha(\vec q,\vec k-\vec q) \delta s_1(\vec k-\vec q) \delta s_1(\vec q)\right)\right]\ .
\eea

Let us explain some of the relevant expressions that appear here. $G(a,\tilde a)$ is the retarded Green's function for the second order  linear differential operator associated with $\delta$ that is obtained after substituting $\theta$ in the second equation of (\ref{eq:master}) with the value obtained from the first, and linearizing. In doing this, it is important to neglect all the terms of order $c_s^2$ because, in our power counting, they count as non-linear terms. This is given by
 \bea \nonumber
&& -a^2{\cal H}^2(a)\d_a^2G(a,\ta)-a \left( 2 {\cal H}^2(a)+ a {\cal H}(a){\cal H}'(a)\right) \d_a G(a,\ta)+3 \frac{ \Omega_m {\cal H}^2_0}{2a} G(a,\ta)=\delta(a-\ta)\ ,\\
&& G(a,\ta)=0 \quad {\rm for } \quad a<\ta\ .
 \eea
For a $\Lambda$CDM cosmology the result can be expressed\footnote{Using, e.g., Mathematica's ``DSolve" function.} as a hypergeometric function, although its form is not particularly illuminating. For all calculations presented here it is sufficient to numerically solve the above differential equation.   This can be easily accomplished by replacing the $\delta(a-\ta)$ on the RHS of the first equation with zero, but starting with the boundary conditions being $G(a,\ta)|_{a=\ta}=0$, and $\frac{\d}{\d a}G(a,\ta)|_{a=\ta}= 1/(\ta {\cal H}(\ta))^2$ . In principle, it is possible to include in the linear equations that determine the Green's function and the growth functions also the higher-order linear terms proportional to $c_s^2$ and $c_v^2$. Doing this amounts to resumming the effect of these pressure and viscous terms. The resulting linear equation can be easely solved numerically, finding for example that the growth factor becomes $k$-dependent, being the more suppressed the higher is the wavenumber~\cite{Enqvist:2010ex}. However, it is not fully consistent to resum these terms without including the relevant loop corrections.

$D(a)$ represents the growth factor at scale-factor-time $a$. In particular, we have written the linear solution as
\be
\delta^{(1)}_l(k,a)=\frac{D(a)}{D(a_0)}\delta s_1(\vec k)\ ,
\ee
with $a_0$ being the present time, and $\delta s_1$ representing a classical stochastic variable with variance equal to the present {\it smoothed} power spectrum
\be
\langle\delta s_1(\vec k)\delta s_1(\vec q)\rangle= (2\pi)^3 \delta^{(3)}(\vk+\vec q) P_{11,l}(k,\Lambda)\ ,
\ee
with $P_{11,l}(k)$ being the smoothing of the linearly computed power spectrum at present time
\be
P_{11,l}(k,\Lambda)=W_\Lambda(k)^2 P_{11,lin}(k)\ .
\ee

A very useful simplification is due to the fact the growth factor and the Green's function are $k$-independent. This is due to the fact that at linear level we can neglect the pressure and viscosity terms that would otherwise induce a $k$-dependence. Because of this, the convolution integrals that would couple time integration and momentum integration nicely split into separate time integrals and momentum integrals that can be simply performed separately. We have tried to underline this in (\ref{eq:delta2}) by adding suitable parenthesis.
Iterating, we obtain the solution for $\delta$ at cubic order $\delta^{(3)}$. For brevity, we report it in App.~\ref{app:delta3}. Notice that in the terms in $\delta^{(2)}$ (and $\delta^{(3)}$) we have neglected the contribution from the pressure and the viscosity, which count as third order terms. They give 
\bea
\label{eq:delta3}
&&\delta^{(3)}_{l,\,c_{\rm\,comb}^2}(\vec k,a)=-\frac{k^2}{D(a_0)}\int^a_0 d\ta\; G(a,\ta)\, \bar c_{\rm\,comb}^2(\ta)\,  D(\ta)\, \delta s_1(\vec k)\ ,
\eea
where $c^2_{\rm\,comb}$ is given by
\be
c_{\rm\,comb}^2(a)=c_s^2(a)+a \frac{D'(a)}{D(a)} c_v^2(a)\ ,
\ee
and it is the combination that is relevant at one-loop order. For the terms multiplying $c_s^2$ and $c_v^2$ in the second equation of (\ref{eq:master}), we can substitute the linear relation 
\be
\theta_l^{(1)}(a,\vec k)=-a {\cal H}\d_a\delta^{(1)}_l(a,\vec k)=-a{\cal H} \frac{D'(a)}{D(a)} \delta^{(1)}_l(a,\vec k)\ .
\ee

Notice that $\delta^{(2)}$ and $\delta^{(3)}$ are the same as the standard ones used in perturbation theory, with just two differences. The first is the important smoothing of the sourcing power spectrum. This makes the convolution integral that we are going to perform next rapidly converging, but also $\Lambda$ dependent.  The second difference is of a more technical nature, and relies on the fact that in standard perturbation theory the time dependence of the non-linear solution is approximated by the linear growth factor $D$ elevated to the power 2 and 3 for $\delta^{(2)}$ and $\delta^{(3)}$, while the momentum dependence is approximated to be the same momentum dependence as in standard EdS universe. This procedure is exact in EdS, but not so in other space times. Some studies~\cite{Takahashi:2008yk} (see also~\cite{martel:1991,Bernardeau:1993qu,Scoccimarro:1997st}) have checked that this is correct up to percent level on the full power spectrum. Since however percent accuracy is the target of next generation experiments, we decide to perform the correct computation, which is not so very complicated to set up in any case. For the purpose of comparing with the literature and to gain familiarity with the EFT setup with simpler formulas, we provide results obtained with this approximate treatment of the perturbed solutions in App.~\ref{app:approx_perturbation_theory}.

\subsubsection{Diagrams}
By contracting the non-linear expression we obtain the non-linear corrections. There are three diagrams at order $\delta^4_l$. After including the linear contribution, we have
\be
\langle\delta_l(\vec k,a_0)\delta_l(\vec q,a_0)\rangle=(2\pi)^3\delta^{(3)}(\vec k+\vec q)\left(P_{11}(k,a_0)+P_{22}(k,a_0)+P_{13}(k,a_0)+P_{13,\;c^2_{\rm\,comb}}(k,a_0)\right)
\ee
with
\bea
&& P_{11}(k,a_0)=\langle\delta^{(1)}(\vk,a_0)\delta^{(1)}(\vkp,a_0)\rangle'\ ,\\ \nonumber
&& P_{22}(k,a_0)=\langle\delta^{(2)}_l(\vk,a_0)\delta^{(2)}_l(\vkp,a_0)\rangle'\ , \\ \nonumber
&& P_{13}(k,a_0)=2\langle\delta^{(3)}_l(\vk,a_0)\delta^{(1)}_l(\vkp,a_0)\rangle'\ ,\\ \nonumber
&& P_{13,\;c^2_{\rm\,comb}}(k,a_0)=2\langle\delta^{(3)}_{l,\,c_{\rm\,comb}^2}(\vk,a_0)\delta^{(1)}_l(\vkp,a_0)\rangle'\ ,
\eea
where the $\langle\ldots\rangle'$ means that we have removed a factor of $(2\pi)^3\delta^{(3)}(\vec k+\vec q)$ from the expectation value. $P_{11}$ represents the {\it unsmoothed} linear power spectrum, as the linear theory does not need to be regularized. The term $P_{13,\;c^2_{\rm\,comb}}$ is supposed to remove the $\Lambda$ dependence that comes from $P_{13}$. It is a counterterm diagram. Strictly speaking, we would need a counterterm diagram also from $P_{22}$, which is provided by the two-point function of the stochastic source $\Delta J^i$ in~(\ref{eq:all_equations}). As we discussed, this term is supposed to count as a $\delta_l^5$ term, and therefore we neglect it. This means that the $\Lambda$ dependence associated with $P_{22}$ is very weak at this order in the calculation.  The full stochastic term will be included in a following paper~\cite{foreman}.

\begin{figure}[h!]
\begin{center}
\includegraphics[width=8.2cm]{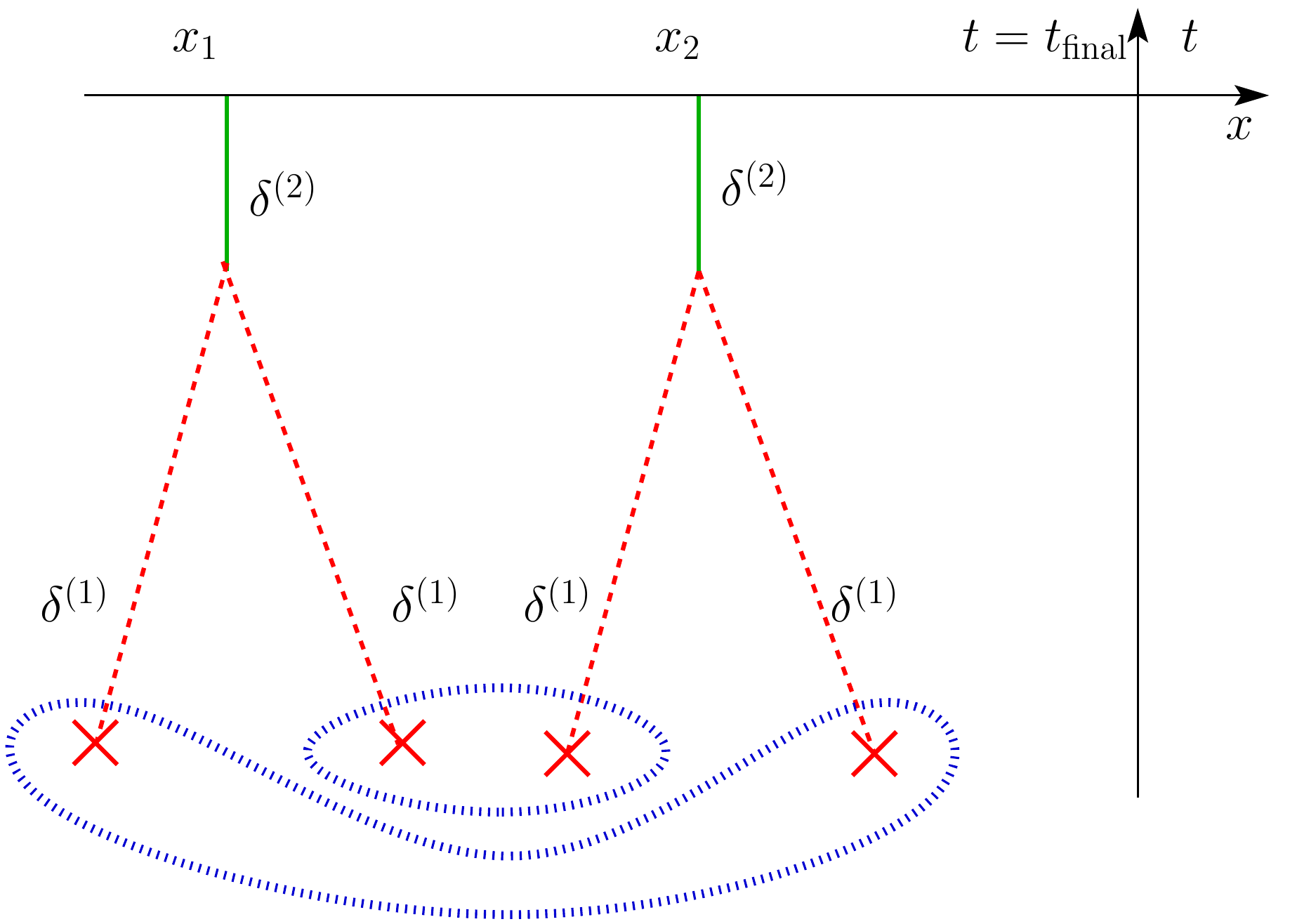}
\includegraphics[width=8.2cm]{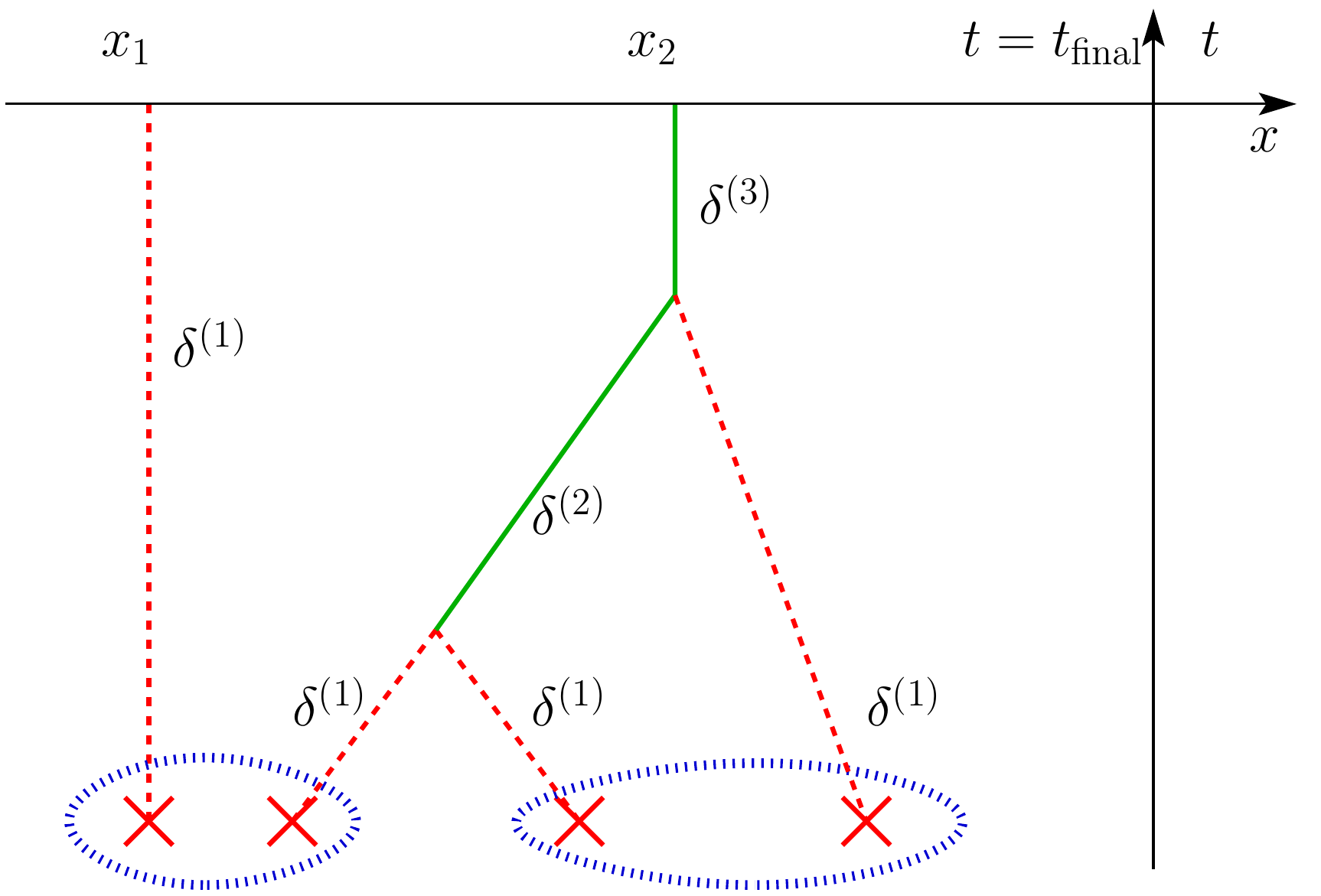}
\includegraphics[width=8.2cm]{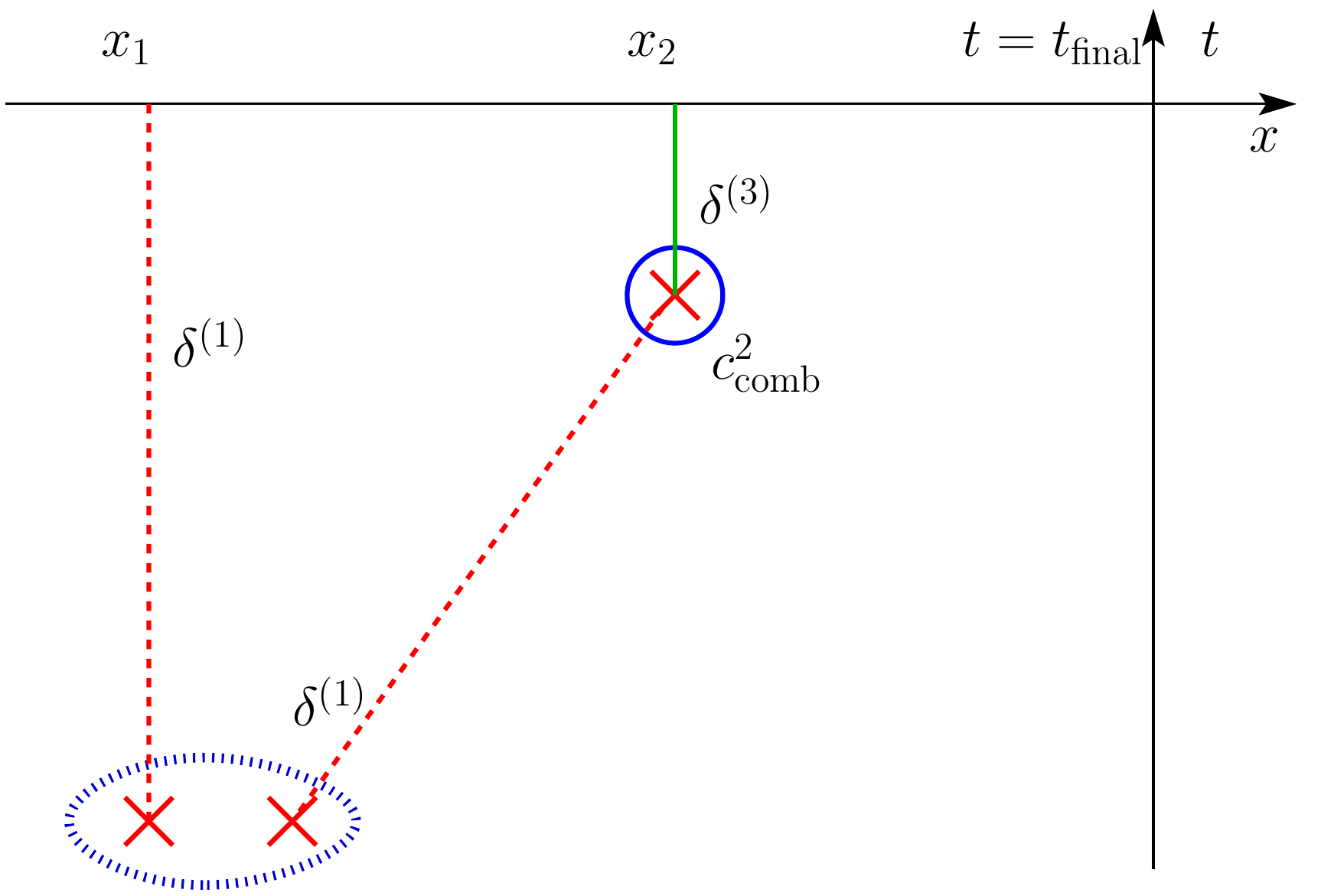}
\caption{\label{fig:diagrams} \small\it Diagrammatic representation of $P_{22}$ (top left),  $P_{13}$ (top right), and $P_{13,\; c^2_{\rm comb}}$ (bottom). Continuous green lines represent Green's functions, red dashed lines represent free fields, and red crosses circled by a dotted blue line represent correlation among free fields.}
\end{center}
\end{figure}

The expressions for $P_{22},\;P_{13},\; P_{13,\;c^2_{\rm\,comb}}$ are given by
\bea\label{eq:P22} \nonumber
&&P_{22}(k,a_0)=\frac{k^3}{16\pi^2 D(a_0)^4} \int dq\, d(\cos\theta)\frac{1}{\left(k^2-2\cos(\theta) k\, q+q^2\right)^2}P_{11,l}(q,\Lambda)P_{11,l}(|\vkkp|,\Lambda) \\ \nonumber
&&\qquad\left[\left(\int_0^{a_0}d\ta\;\ta^2 G(a_0,\ta)\, {\cal H}^2(\ta)D'(\ta)^2\right)4\cos(\theta)(k-\cos(\theta) q)\right.\\ \nonumber
&&\qquad\qquad\qquad\qquad\left.+3{\cal H}_0^2 \Omega_m\left(\int_0^{a_0}d\ta\;G(a_0,\ta)\,\frac{D(\ta)^2}{\ta}\right)\left(\cos(\theta) \left(k-2 \cos(\theta) q\right)+q\right)\right]\times\\ \nonumber
&&\qquad\left[\left(\int_0^{a_0}d\ta\;\ta^2 G(a_0,\ta)\, {\cal H}^2(\ta)D'(\ta)^2\right)\left(3 k^2 \cos(\theta)-k\, q\left(4\cos(\theta)^2+1\right)+2\cos(\theta) q^2\right)\right.\\
&&\qquad\qquad\qquad\qquad\left.+3{\cal H}_0^2 \Omega_m \left(\int_0^{a_0}d\ta\;G(a_0,\ta)\,\frac{D(\ta)^2}{\ta}\right)\left(k^2-2\cos(\theta) k\, q+q^2\right) \right]\ ,
\eea
where $\cos(\theta)=\vk\cdot\vkp/(k\, q)$;
\bea\label{eq:P13}
&&P_{13}(k,a_0)=-\frac{2~ k^3}{96 (2\pi)^2 D(a_0)^3}P_{11,l}(k,\Lambda)\\ \nonumber
&&\qquad\int_0^\infty \frac{dr}{r^3} \left[12 r^7 {\cal D}_4-24 r  {\cal D}_5+4 r^3\left(16  {\cal D}_1+8  {\cal D}_2+4  {\cal D}_3-3 {\cal D}_4+24  {\cal D}_5\right)\right.\\ \nonumber 
&&\qquad\qquad\qquad\left. +8 r^5 \left(4 {\cal D}_2+2 {\cal D}_3-6 {\cal D}_4+3 {\cal D}_5-4 {\cal D}_6\right)\right.\\ \nonumber
&&\qquad \qquad\qquad \left.+3 \left(r^2-1\right)^3\left(r^2  {\cal D}_4+2 {\cal D}_5\right)\log\left(\frac{(1-r)^2}{(1+r)^2}\right)\right] P_{11,l}(k\, r,\Lambda)\ ,
\eea
where ${\cal D}_{1,\ldots,6}$ are given in App.~\ref{app:delta3},
and finally
\bea\label{eq:P13cs}
 P_{13,\;c^2_{\rm\,comb}}(k,a_0)=-2\frac{k^2}{D(a_0)}\int^{a_0}_0 d\ta\; G(a_0,\ta)\, c_{\rm\,comb}^2(\ta)\,  D(\ta)\,P_{11,l}(k,\Lambda)\ .
\eea
The convolution integrals in $P_{22}$ and $P_{13}$ are the sign that these are one loop diagrams. $ P_{13,\;c^2_{\rm\,comb}}$ does not have a convolution integral as it is a one-loop counterterm.  The diagrams are pictorially represented in Fig.~\ref{fig:diagrams}. Since $\alpha(\vk,-\vk)=\beta(\vk,-\vk)=0$, there are no non-1PI diagrams. Notice that in $P_{13}$ and $P_{22}$ we have already carried out at least a part of the angular integration.  The sign of the $ P_{13,\;c^2_{\rm\,comb}}$ in (\ref{eq:P13cs}) is also quite intuitive. For positive $c_s$ or $c_v$, the gravitational collapse is slowed down, and so this contribution tends to decrease the gravitational collapse.

\subsubsection{Cutoff-(in)dependence}

Each diagram is dependent on the cutoff $\Lambda$: $P_{22}$ and $P_{13}$ through the smoothed linear power spectrum, while $P_{13,\;c^2_{\rm\,comb}}$ depends on $\Lambda$ through $c^2_{\rm\,comb}$, which is $\Lambda$ dependent because it arises from integrating out the short distance fluctuations. The $\Lambda$ dependence of  $P_{13}$ is to be cancelled by $P_{13,\;c^2_{\rm\,comb}}$, while the one of $P_{22}$ from the stochastic fluctuations.

\begin{figure}[h!]
\begin{center}
\includegraphics[width=10cm]{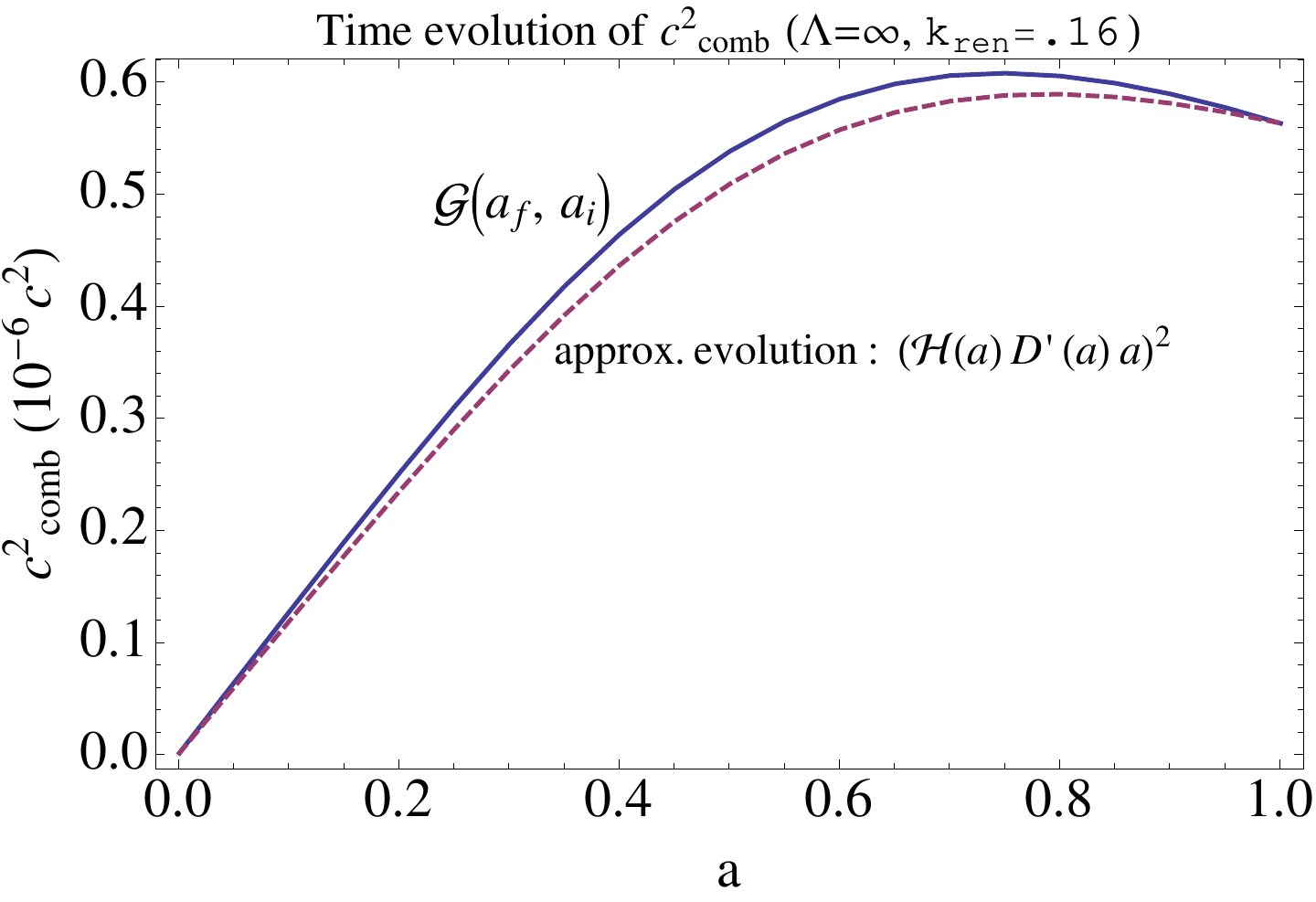}
\caption{\label{fig:csCombTimeDep} \small\it Time dependence of $c^2_{\rm\,comb}$ as inferred using the correct time dependence from $P_{13}$ and instead using the approximate time dependence derived from the growth functions (see App.~\ref{app:approx_perturbation_theory}). Starting from very early times, we see that $c^2_{\rm\,comb}$ grows as a functions of time, peaks at about $a\simeq 0.7$, and then decreases near the present epoch, probably as due to the onset of dark energy. $c^2_{\rm\,comb}$ is positive, implying that this term tends to slow down the collapse of structures.}
\end{center}
\end{figure}

Consider the sum of the $P_{13}$ and $P_{13,\;c^2_{\rm\,comb}}$ terms. In order for this sum to be $\Lambda$ independent, both must have the same $k$-dependence in the relevant regime. By inspection of (\ref{eq:P13cs}), we see that $P_{13,\;c^2_{\rm\,comb}}$ goes as $k^2$, which implies that $P_{13}$ should behave in the same way in this regime. This is in fact the case, as can be readily verified by taking the $k\rightarrow 0$ limit of~(\ref{eq:P13}). In particular, we can define a $\Lambda$-independent renormalized parameter $c_{\rm\,comb; ren.}$ defined at a renormalization scale $k_{\rm ren.}$ and a $\Lambda$-dependent counterterm parameter $c^2_{\rm\,comb; ctr.}(\Lambda)$ such that
\be
c_{\rm\,comb}^2(a,\Lambda)=c_{\rm\,comb; ren.}^2(a,k_{\rm ren.})+c_{\rm\,comb; ctr.}^2(a,\Lambda)\ .
\ee
$c^2_{\rm\,comb; ctr.}(a,\Lambda)$ must have the same time and $\Lambda$-dependence of $P_{13}$, while $c_{\rm\,comb; ren.}(a,k_{\rm ren.})$ is determined by matching to simulations or to observations at a specific $k=k_{\rm ren.}$. The time dependence of $c^2_{\rm\,comb; ren.}(a)$ could in general be different from the one of $c^2_{\rm\,comb; ctr.}(a,\Lambda)$. However, for the purposes of this paper, we can approximate them to be equal, and we will check this approximation in a forthcoming paper~\cite{foreman}. We can therefore extract the time dependence of $c^2_{\rm\,comb}(a,\Lambda)$ from the $k\rightarrow 0$ limit of $P_{13}$. We obtain
\bea
c_{\rm\,comb; ctr.}^2(a,\Lambda)=c_{\rm\,comb; ctr.}^2(a_0,\Lambda)\frac{{\cal}D_{c^2_{\rm comb}}(a)}{{\cal}D_{c^2_{\rm comb}}(a_0)}\ ,
\eea
where
\bea
&&{\cal}D_{c^2_{\rm comb}}(a)=\frac{1}{a }\left[\left(63\,{\cal H}_0^4\Omega_m^2 {\cal F}_1(a)+12\,{\cal H}_0^2\Omega_m{\cal F}_2(a)+52\, a^3 {\cal H}^2(a)D'(a)^2\right)\right.\\ \nonumber
&&\qquad\qquad\qquad\quad\left.-4\, a^3{\cal H}^2(a) \frac{D'(a)}{D(a)}\left(27\, {\cal H}_0^2\Omega_m {\cal F}_3(a)+28\, {\cal F}_4(a)\right)\right]\ .
\eea
and
\bea
&&{\cal F}_1(a)=\int_0^a d\ta\; G(a,\ta)\;\frac{D(\ta)^2}{\ta}\ , \qquad\qquad {\cal F}_2(a)=\int_0^a d\ta\; G(a,\ta)\;\ta^2 D'(\ta)^2 {\cal H}^2(\ta)\ , \\ \nonumber
&&{\cal F}_3(a)=\int_0^a d\ta\; \d_aG(a,\ta)\; \frac{D(\ta)^2}{\ta}\ , \quad\qquad {\cal F}_4(a)=\int_0^a d\ta\; \d_aG(a,\ta)\;\ta^2 D'(\ta)^2 {\cal H}^2(\ta)\ . 
\eea
A plot of the time dependence of the speed of sound is given in Fig.~\ref{fig:csCombTimeDep}. 
 
We determine the value of $c^2_{\rm\,comb; ctr.}(a_0,\Lambda)$ in two different independent ways: one involving fitting to observation of the power-spectrum derived from simulations, and the other involving direct measurement from simulation. In the first, simplest, way, we determine $c^2_{\rm\,comb; ctr.}(a_0,\Lambda=\infty)$ by matching the one-loop EFT power spectrum at $k=k_{\rm ren.}$  to the power-spectrum extracted from simulations (or directly from precise observations in the future!).  We can do this at various values of $\Lambda$, but we take the $\Lambda\rightarrow\infty$ limit in order to drive to zero any effect from higher derivative terms down by powers of $k/\Lambda$. At this point, we can derive $c^2_{\rm\,comb}(a_0,\Lambda\neq\infty)$ by running the value at $\Lambda=\infty$ down to a finite $\Lambda$. The formula is given by
\begin{multline}\label{eq:cs_running}
c^2_{\rm\,comb}(a_0,\Lambda\neq\infty)\,= \, c^2_{\rm\,comb}(a_0,\Lambda=\infty) \,+
\lim_{k_{\rm ext}\to 0} \Bigg[\Big({P_{13}(k_{\rm ext},a_0,\Lambda=\infty)-P_{13}(k_{\rm ext},a_0,\Lambda)} \Big) \times \\
\left({-2\frac{k_{\rm ext}^2}{D(a_0)}\int^{a_0}_0 d\ta\; G(a_0,\ta)\, \frac{{\cal}D_{c^2_{\rm comb}}(\ta)}{{\cal}D_{c^2_{\rm comb}}(a_0)}\,  D(\ta)\,P_{11,l}(k_{\rm ext},\Lambda)}\right)^{-1}\Bigg]\ .
\end{multline}

\begin{figure}[h!]
\begin{center}
\includegraphics[width=11cm]{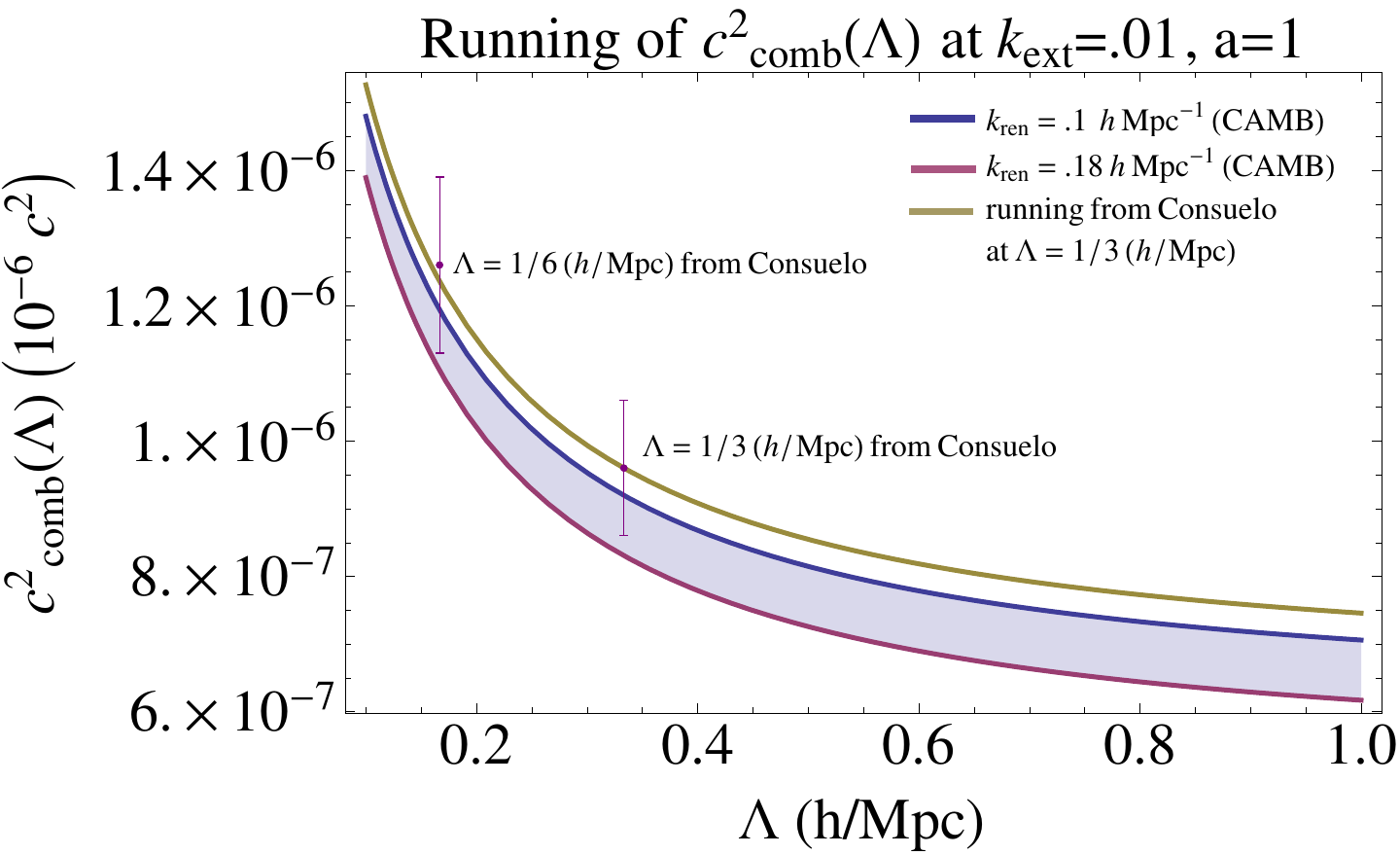}
\caption{\label{fig:cscomb_running} \small\it Running of $c_{\rm\,comb}^2$ as a function of $\Lambda$. The purple band contains the region for the values of $c_{\rm\,comb}^2$ as inferred from matching with the non-linear power spectrum from CAMB at the renormalization scale $k=0.1h$ Mpc$^{-1}$ and $k=0.18h$ Mpc$^{-1}$. The dependence on the renormalization scale is a measure of the importance of higher loops. We see that as $\Lambda\to\infty$,  $c_{\rm\,comb}^2$ decreases as more and more modes are included within the regime of validity of the EFT. However, the fact that as $\Lambda=\infty$, $c_{\rm\,comb}^2\neq 0$ is an indication of the fact that the fundamental theory is not described by a pressureless ideal fluid, but by indeed freely streaming dark matter particles. Data points with  $1\,\sigma$ error bars represent the value obtained from $N$-body numerical simulations using the methods described in Sec.~\ref{sec:simulation_results} using two different smoothing lengths $\Lambda^{-1}$. Given the error bars from numerical evaluation, the measured values are in remarkable agreement with what inferred from renormalizing using the power spectrum. }
\end{center}
\end{figure}

The limit $k_{\rm ext}\to 0$ is necessary in order to suppress higher derivative terms down by powers of $k_{\rm ext}/\Lambda$.
The running of  $c^2_{\rm\,comb}$ is plotted in Fig.~\ref{fig:cscomb_running}. $c^2_{\rm\,comb}(a_0,\Lambda=\infty)\simeq 6.2\times 10^{-7}$ and is the value obtained by fitting to data using $k_{\rm ren.}=0.16\, h$  Mpc$^{-1}$~\footnote{It is somewhat interesting to notice that by using $c_{sv}\sim 10^{-7}$ we find a shear viscosity of order $\eta\sim20$~Pa~s, in SI units. This value is very similar to that of some everyday items such as chocolate syrup!}. We see that as $\Lambda$ decreases, we integrate out more and more modes, and $c^2_{\rm\,comb}$ grows. We see that the result  matches with the one obtained by the second method for measuring $c^2_{\rm\,comb}$, that is by using $N$-body simulation to extract directly the correlations in~(\ref{eq:parameters}). We perform measurements at $\Lambda=1/3\,h$ Mpc$^{-1}$ and $\Lambda=1/6\,h$ Mpc$^{-1}$, and we see that the match is extremely good. We describe more precisely how these measurements are derived in Sec.~\ref{sec:simulation_results}. We take this as an extremely promising indication of the strength of our approach. 

\begin{figure}[h!]
\begin{center}
\includegraphics[width=9cm]{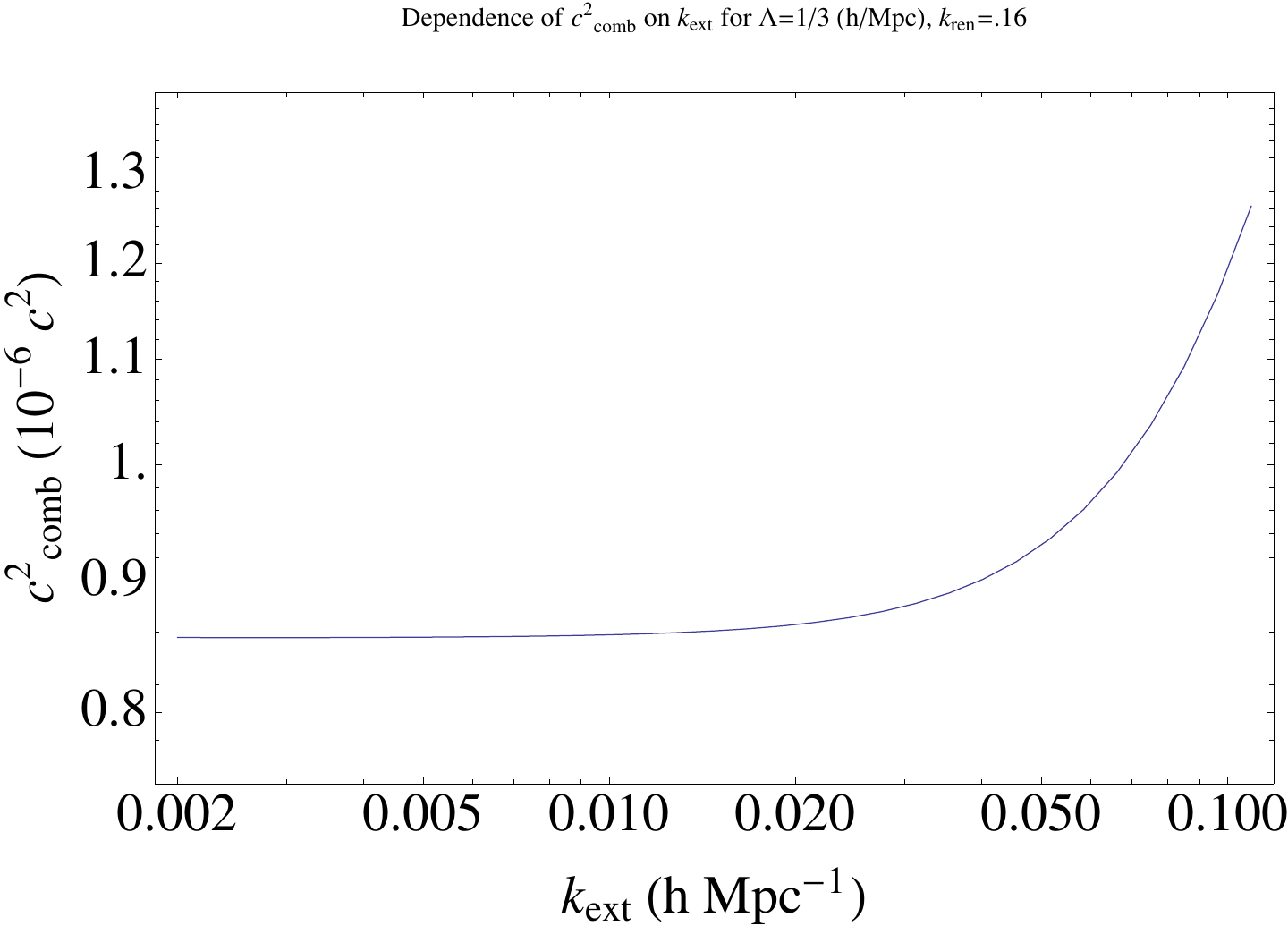}
\caption{\label{fig:higher_derivative_cs} \small\it  In this plot we present to values obtained for $c^2_{\rm\,comb}$ as a function of the external momentum used in (\ref{eq:cs_running}). We see that only as $k_{\rm ext}\to0$, $c^2_{\rm\,comb}$ becomes $k_{\rm ext}$ independent. This is so because at high $k$ any higher derivative terms suppressed by powers of $k/k_{NL}$ are important.}
\end{center}
\end{figure}

In order to elucidate the effect of the higher derivative terms, we plot in Fig.~\ref{fig:higher_derivative_cs} the value of $c^2_{\rm\,comb}(a_0,\Lambda=1/3)$, for various values of the external $k_{\rm ext}$. It is only for vary low $k_{\rm ext}$'s that $c^2_{\rm\,comb}$ becomes $k_{\rm ext.}$ independent.

Finally, there is a third method in which we could have derived $c^2_{\rm\,comb}(a_0,\Lambda)$. By keeping $\Lambda$ finite, we could have fit the analytical results to $N$-body simulations by including higher derivative terms proportional to powers of $k/\Lambda$. Indeed, unless $\Lambda\rightarrow\infty$, the largest of these terms are not negligible and need to be included to get the correct $c^2_{\rm\,comb}(a_0,\Lambda)$. A description of this approach in detail is given in App.~\ref{app:higher-deriv}, and leads to the same results for $c^2_{\rm\,comb}(a_0,\Lambda)$.

\section{Fluid parameters from $N$-body simulations\label{sec:simulation_results}}

\label{Scn:Measurement}


If this language of effective field theory is to be born out, we must be able to take the fundamental theory, integrate away UV effects, and find agreement in terms of the parameters described above.  Fortunately we have, in the form of simulation, exactly those calculations in the fundamental theory.  By smearing the positions of simulated particles with a normalized Gaussian function of width $\Lambda$,  we are able to introduce a soft UV cutoff of order $1/\Lambda$.  For correlations on scales longer then the cutoff we can directly measure $c_{\rm comb}^2$ corroborating the perturbative analysis presented above. 

Specifically we  consider random downsamples of positions and velocities of the {\it Consuleo} simulation\footnote{The simulation parameters are: $\Omega_m = 0.25$, $\Omega_\Lambda = 0.75$, $ h = 0.7$  ($H = 70~{\rm km/s/Mpc}$), $\sigma_8 = 0.8$, and $n_s = 1$ with measurements described taking place at $z=0$.}~\cite{consuelo}  from $2.7\times10^9$ particles downsampled to $1,000,053$ particles distributed over $(420~{\rm Mpc})^3$.   Even this incredibly coarse resolution allows us to measure the following fields $\delta_l$, $\Theta_l$, $\partial^2 \delta_l$,  $\partial^2 \Theta_l$, and $A_s$ to measure $\csc$ to within standard errors of $10$ percent.  While the complexity of the first four fields go linearly in the number of particles, $A_s$ is more expensive. The fact that we can achieve such consistency with such a small number of particles  is not only remarkable, but numerically quite convenient.  We describe the details of the analysis in \app{App:MeasurementDetails}, and here simply provide a summary of the results.

\begin{figure}[h!]
    \centering
        \includegraphics[width=0.49\textwidth]{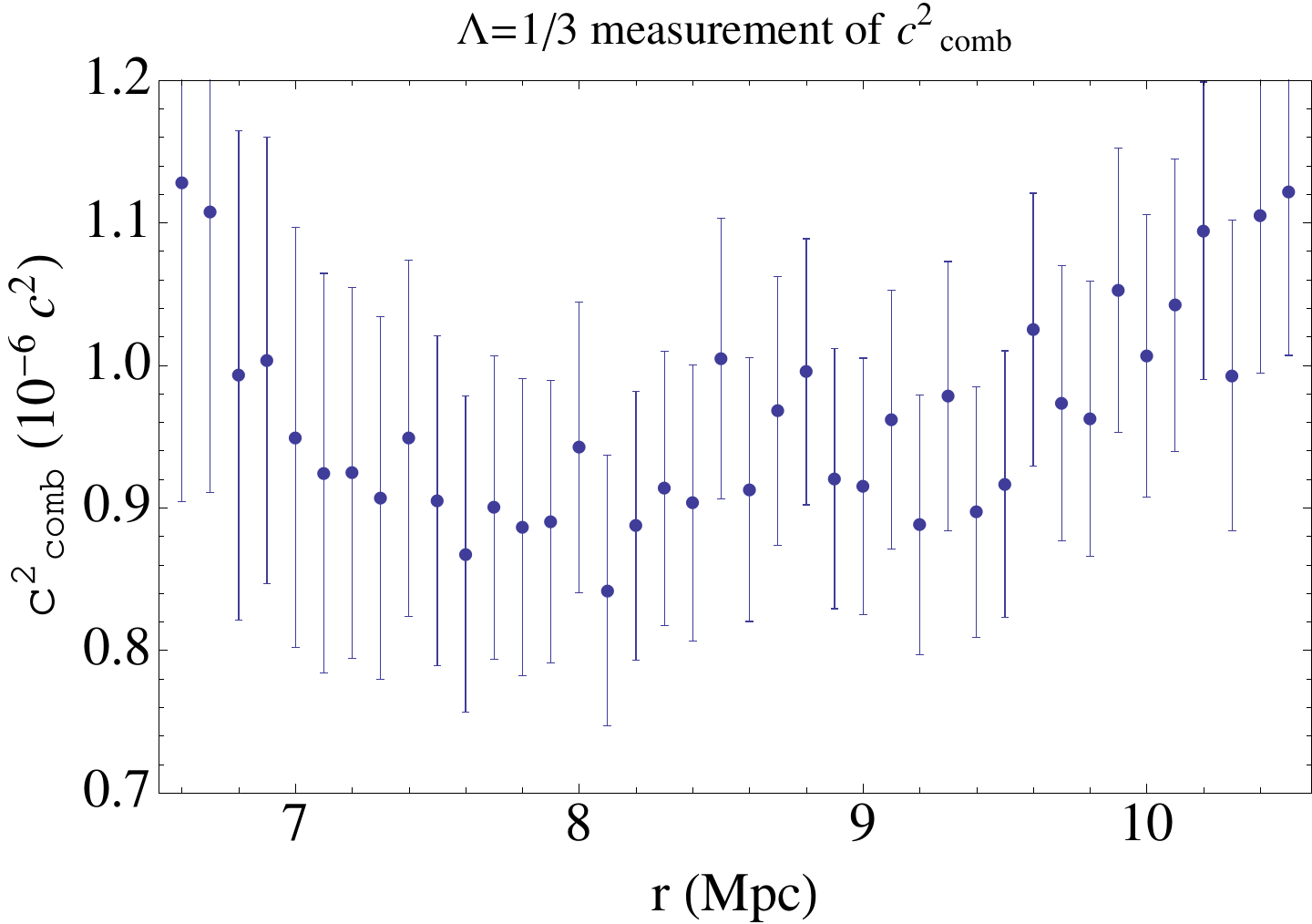}
           \includegraphics[width=0.49\textwidth]{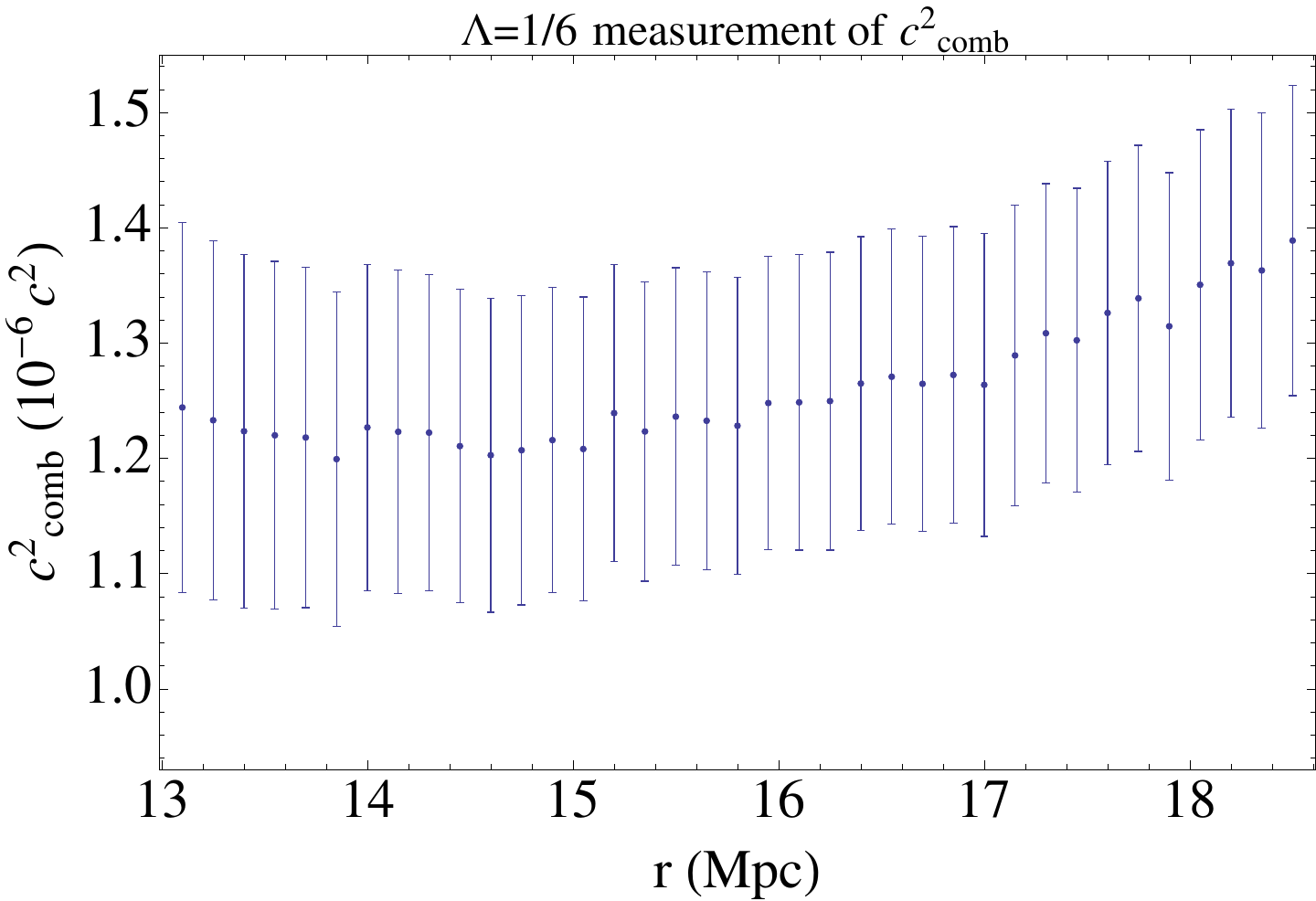}
    \caption{\sl  Measurement of  $\csc$ in the UV with $\Lambda=1/3~  h$ Mpc$^{-1}$ and $\Lambda=1/6~  h$ Mpc$^{-1}$.}
    \label{spatialL3}
\end{figure}

The spatial dependence of measured $\csc$ are plotted for $\Lambda=1/3\,h$~Mpc$^{-1}$ and $\Lambda=1/6\,h$~Mpc$^{-1}$ in  \Fig{spatialL3}.  These regions were chosen to maximize numerical stability as described in \app{app:simulations}.  Fitting a constant to the value of $\csc$ for the displayed values after the UV cutoff gives in units of $c^2$, we obtain
\begin{align}
\csc{(\Lambda=1/3)} &=0.96 \pm .1 \times 10^{-6}\, , \\ \nonumber
\csc{(\Lambda=1/6)} &=1.26 \pm .1 \times 10^{-6}\, .
 \end{align}

The RG flow between the two measured values is consistent with the prediction from perturbation theory as seen in \Fig{fig:cscomb_running}. Furthermore the measured value from Consuelo agrees nicely with that predicted from matching  to the nonlinear CAMB power spectrum.

\section{Results}

\begin{figure}[h!]
\begin{center}
\includegraphics[width=0.79\textwidth]{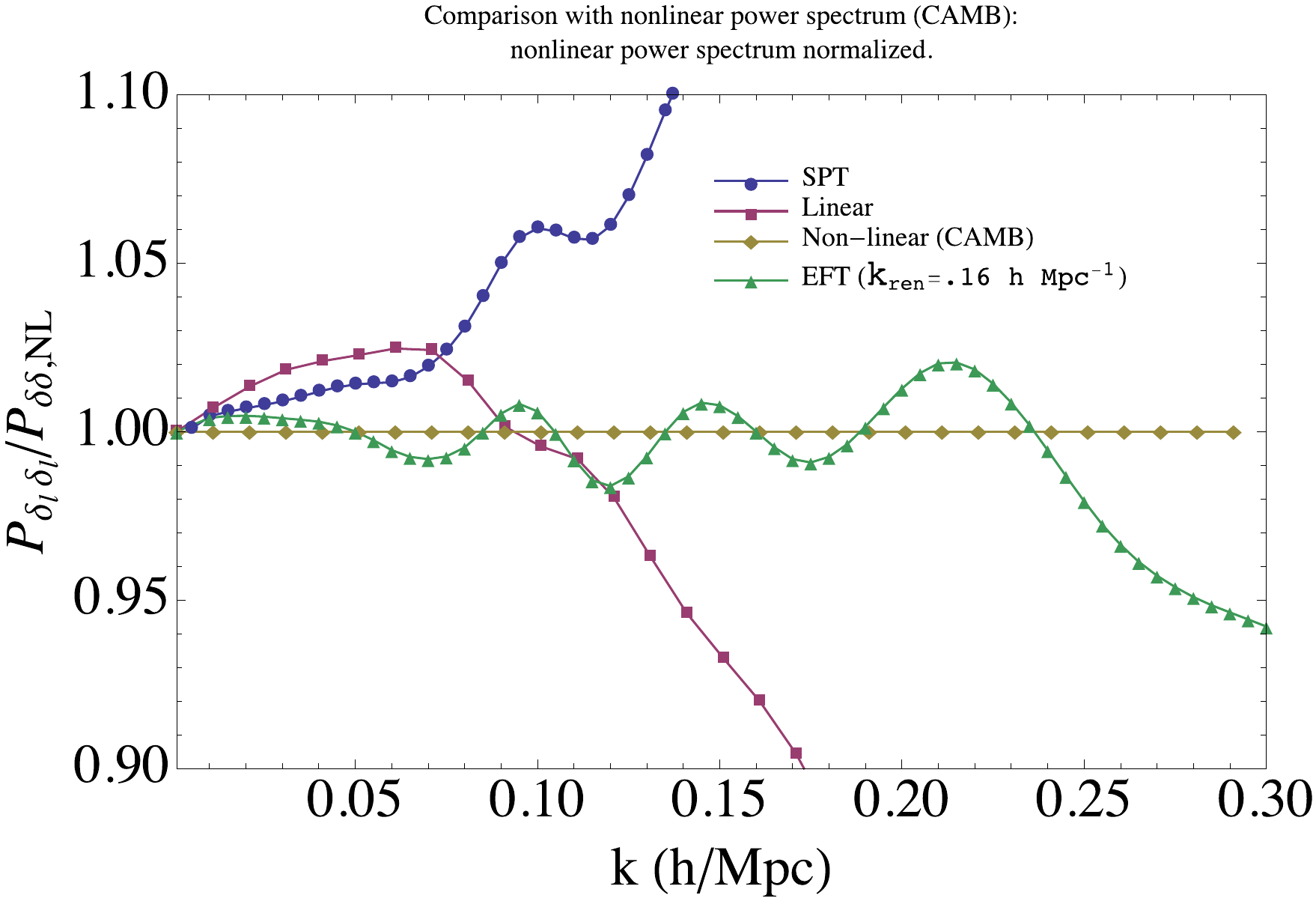}
\includegraphics[width=0.74\textwidth]{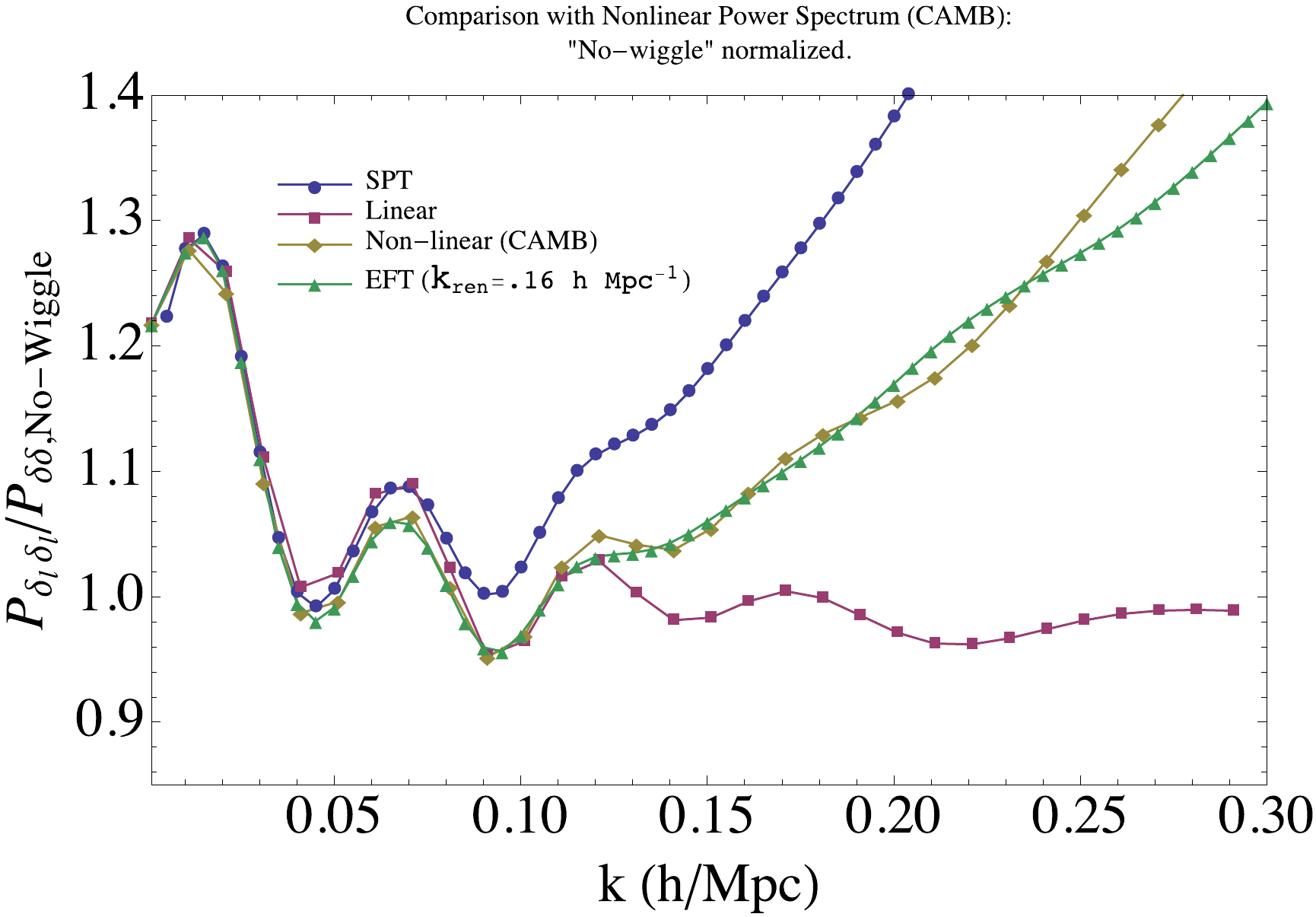}
\caption{\label{fig:EFT_prediction1} \small\it  The order $\delta_l^4$ prediction from our EFT is compared with the CAMB non-linear output in the top, and to the no-wiggle power spectrum in the bottom, as well  with the linear theory and Standard Perturbation Theory (SPT). The results from the EFT agree at percent level with the non-linear theory up to $k\simeq 0.24 h$ Mpc$^{-1}$, when some high scale power seems to be missing. Results should improve already by going to $\delta_l^5$ order. The results are remarkably better than using SPT. The no-wiggle power spectrum we use is given by $P_{\delta\delta,{\rm No-Wiggle}}=5.1\cdot 10^6 q \log ^2(13 q+2 e)/(54\, q^2
  (14 +731/(457 q+1))+\log (13 q+2 e))^2$.}
\end{center}
\end{figure}

The ${\cal{O}}(\delta_l^4)$ result of the computation of the power spectrum with our EFT is presented in Fig.~\ref{fig:EFT_prediction1}. On top, we plot the ratio of the one loop power spectrum compared with the non-linear fit provided by the CAMB software with high precision settings, evaluated with the following cosmological parameters: $\Omega_{\Lambda}=0.75\;,\Omega_m=0.25,\;\Omega_b=0.04,\; h=0.7,\;n_s=1$. The linear power spectrum is also obtained from CAMB with high precision settings. We take these data for all perturbative calculations done in the paper. Often in the literature the results of perturbation theory are plotted as ratio of the perturbation theory result versus a no-wiggle power spectrum. In this way the oscillatory features are still present in the plot, though they come mostly from the linear theory. We give this in the bottom part of the plot.

This plot is obtained after renormalizing the EFT prediction to match the power spectrum at $k_{\rm ren.}=0.16 h\ $Mpc$^{-1}$. In detail, we perform the calculation at several increasing values of $\Lambda$, we choose the value of $c^2_{\rm\,comb}(a_0,\Lambda)$ to match the simulations' non-linear output, and then we extrapolate to $\Lambda\to \infty$. This gives us $c^2_{\rm\,comb}(a_0,\Lambda=\infty)\simeq 6.2\times 10^{-7}$. Notice that naive considerations of virialization gave an estimated value for $c^2_{\rm\,comb}$ of order $10^{-5}$~\cite{Baumann:2010tm}. The obtained smaller numerical value fits well with the decrease in the transfer functions for wavenumbers that are higher than the equality scale.

The result for the power spectrum agrees at percent level with the CAMB non-linear fit up to $k\simeq0.24 h$ Mpc$^{-1}$, where the EFT prediction begins to be smaller than the $N$-body simulation result. Results obtained with the approximate time dependence described in App.~\ref{app:approx_perturbation_theory} are close to these ones, at percent level. As discussed in App.~\ref{app:approx_perturbation_theory}, it is potentially dangerous to trust this approximation at high $k$'s for percent level precision, and luckily it is not very hard at all to perform the correct perturbation theory. We expect that the inclusion of the stochastic pressure and of higher order diagrams should improve the fit in the UV, possibly allowing us to fit the simulations to even higher $k$'s. Notice how the counterterm $P_{13,\,c_{\rm comb}^2}$ decreases the power spectrum, compensating for the overshooting of SPT. It is difficult to interpret the percent disagreement that we have at moderate slow scales such as at $k\simeq 0.12 h$ Mpc$^{-1}$. At face value, it looks like that the computed power spectrum presents  oscillations that are too large. These disagreements might be improved with the inclusion of higher order terms, or it might even be that at this level of precision, the results from the CAMB non-linear fit or from the $N$-body simulations might not be precise enough. These same improvements should reduce also the dependence on the renormalization scale: by changing the renormalization scale our results change by about 2\%, at $k\sim0.24 h$~Mpc$^{-1}$. This dependence can be taken as a measure of the contribution from higher order terms.


\section{Conclusions}

Large scale structure surveys have the potential of becoming the next leading observational window on the physics of the early universe, potentially greatly improving what we are already learning from the CMB. Large Scale Structure physics is however much more complicated than the CMB due to the presence of large matter clustering at small scales. Since at non-linear level different scales are coupled, these non-linearities affect even large scale perturbations that are mildly non linear and so potentially treatable in a perturbative matter. In this paper we have developed the effective field theory of cosmological large scale structures in order to achieve a reliable predictability. Calculations in the effective theory are performed in $k/k_{NL}$. The effective field theory is a cosmological fluid description for cold dark matter, and by extension all matter including baryons which trace the dark matter.
The microphysical description is in terms of a classical gas of point particles, which we have smoothed at the level of the Boltzmann equation. We have exhibited and computed the various couplings that appear in the effective field theory, namely pressure and viscosity by matching to $N$-body simulations, finding $c_s^2\sim 10^{-6}\,c^2$, etc.\,. We have developed the perturbative expansion for the power spectrum, which we have carried out at the ${\cal O}(\delta_l^4)$. The fluid parameters arise from UV modes and alter standard perturbation theory. We have found that the corrections lead to a power spectrum in percent agreement with the full nonlinear spectrum as obtained by CAMB up to $k\simeq 0.24 h$~Mpc$^{-1}$.

It is a peculiar coincidence that $k\simeq0.24 h$ Mpc$^{-1}$ is also the maximum $k$ at which the popular technique of Renormalized Perturbation Theory (RPT)~\cite{Crocce:2005xy} works. While RPT is a very nice technique to compute non-linear corrections to the power spectrum, we stress that our approach is different at a qualitative and a quantitative level. At a qualitative level, RPT tries to solve, as exactly as possible, non-linear equations for a pressureless ideal fluid. In our approach, instead, we try to solve non-linear equations for a different fluid. This has quantitative effects, as it is shown from the fact that as $\Lambda\to\infty$ our effective parameters like $c_{\rm comb}^2$ do not vanish. What is more important about our EFT is that in can be improved. By performing higher order computations and by adding suitable counterterms, in principle arbitrary precision for reconstructing the power spectrum, or indeed any dark matter observables, can be achieved by going to a sufficiently high order in perturbation theory, on scales $k\lesssim k_{NL}$. Techniques such as RPT or the renormalization group approach~\cite{Matarrese:2007aj} for example are still very nice techniques to perturbatively solve some non-linear equations, resumming many diagrams. It would be interesting to apply those techniques to solve the equations of our EFT. We leave this to future work.

The effective field theory approach to large scale structure formation is complimentary to $N$-body simulations by providing an elegant fluid description. This  provides intuition for various nonlinear effects, as well as providing computational efficiency, since the numerics required to measure the fluid parameters are expected to be computationally less  expensive than a full scale simulation. Indeed we are able to achieve excellent agreement with the small down-sample of the full simulation we examined here.  Of course, since the couplings are  UV sensitive, it still requires the use of some form of $N$-body simulation to fix the physical parameters, either by matching to the stress-tensor directly or to observables. But this matching is only for a small number of physical parameters at some scale and then the constructed field theory is predictive at other scales.

There are several possible extensions of this work. A first extension is to go beyond the one-loop order to two-loop, or higher. This will require the measurement of several new parameters that will enter the effective stress-tensor at higher order, including its stochastic terms. Another extension is to compute the velocity fields and  to include the small but finite contributions from vorticity, or to compute higher order $N$-point functions, which can probe non-Gaussianity.
Finally, another extension is to consider different cosmologies; in this work  we have presented results on dark energy in the form of a cosmological constant. But one could equally consider other models for dark energy. This would presumably alter the value of the fluid parameters in a way that could be measured either from new simulations or determined by observation. Hopefully, our effective field theory for large scale structures will help us use large scale structure surveys to uncover the physics of the beginning of the universe.

\section*{Acknowledgments}
We thank T.~Abel, A.~Arvanitaki, T.~Baldauf, D.~Baumann, P.~Behroozi, S.~Dubovsky, S.~Foreman, S.~Kachru, A.~Nicolis, U.~Seljak, G.~Villadoro, R.~Wechsler and M.~Zaldarriaga for many useful conversations. We gratefully acknowledge Academic Technology Services at UCLA for computer support.  We are also thankful to M. Busha and R. Wechsler for providing us with data on the dark
matter distribution from the Consuelo simulation, which was run using the orange cluster at SLAC by M. Busha as part of the LASDAMAS project.
J.J.M.C. is supported by the Stanford Institute for Theoretical Physics and the NSF grant no. PHY-0756174.  M.P.H. is supported by NSF grant PHY-0756174 and a Kavli Fellowship. L.S. is supported by DOE Early Career Award DE-FG02-12ER41854 and by NSF grant PHY-1068380.

\section*{Appendix}
\begin{appendix}

\section{UV Stress Tensor directly from Short Modes\label{app:short_modes}}

It is useful to write down the expression (\ref{eq:short_stress}) directly in terms of short wavelength fluctuations. In this Appendix we provide these expressions. We define
\bea
\sigma_s^{ij}\amp\equiv\amp m^{-1}a^{-5}\!\int\! d^3{\bf p}\,(p^i-p^i_l({\bf x}))(p^j-p^j_l({\bf x}))f({\bf x},{\bf p}) \\
\amp=\amp \sum_n {m\over a^3}(v_n^i-v_l^i({\bf x}_n))(v_n^j-v_l^j({\bf x}_n))\,\delta^{(3)}_D({\bf x}-{\bf x}_n)\ , \,\,\,\,\,\,\,\,\,\,\, \\
\phi_{s,n}\amp\equiv\amp \phi_n-\phi_{l,n}\ , \\
\partial_i\phi_s\amp=\amp \sum_n\partial_i\phi_{s,n}\ , \\
w^{ij}_s \amp\equiv\amp 
\partial_{i}\phi_s\,\partial_{j}\phi_s-\sum_n\partial_i\phi_{s,n}\,\partial_j\phi_{s,n}\ ,
\eea
where $p_l^i({\bf x})\equiv m \, a \, v_l^i({\bf x})$.
Note that $\sigma^{ij}_s\neq\sigma^{ij}-\sigma_l^{ij}$, but they are related as follows
\bea
\sigma_l^{ij}\amp=\amp\left[\sigma_s^{ij}\right]_\Lambda+\left[\rho_mv_l^iv_l^j\right]_\Lambda+
\left[v_l^i(\pi^j-\rho_mv_l^j)+v_l^j(\pi^i-\rho_mv_l^i)\right]_\Lambda\ .
\eea
The second term is approximately $\rho_lv_l^iv_l^j$ (so it approximately cancels with $-\rho_lv_l^iv_l^j$ in $\kappa_l^{ij}$) and the final term is small (as it is an overlap between short and long modes). 
Following the methods of \cite{Baumann:2010tm} we obtain
\bea
\kappa_l^{ij}=\left[\sigma_s^{ij}\right]_\Lambda+{\rho_l\partial_kv_l^i\partial_kv_l^j\over\Lambda^2}+\mathcal{O}\left(1\over\Lambda^4\right)\ .
\eea
Similarly, one can prove that $\Phi^{ij}_l$ satisfies
\bea
\Phi^{ij}_l \amp=\amp -{[w^{kk}_s]_\Lambda\delta^{ij}-2[w^{ij}_s]_\Lambda\over 8\pi G\,a^2}+\nonumber\\
\amp\amp \!\!\!\!\!{\partial_m\partial_k\phi_l\partial_m\partial_k\phi_l\delta^{ij}-2\partial_m\partial_i\phi_l\partial_m\partial_j\phi_l\over 8\pi G\,a^2\Lambda^2}+\mathcal{O}\!\left(1\over\Lambda^4\right)\ . \,\,\,\,\,\,\,\,\,\,
\eea

So altogether we obtain the effective stress-tensor
\bea
\left[\tau^{ij}\right]_\Lambda \amp=\amp\left[\tau_s^{ij}\right]_\Lambda+\left[\tau^{ij}\right]^{\partial^2}\ ,
\eea
where
\bea
\left[\tau_s^{ij}\right]_\Lambda\amp=\amp \left[\sigma_s^{ij}\right]_\Lambda
-{[w^{kk}_s]_\Lambda\delta^{ij}-2[w^{ij}_s]_\Lambda\over 8\pi G\,a^2} \ , \\
\left[\tau^{ij}\right]^{\partial^2}\amp=\amp {\rho_l\partial_kv_l^i\partial_kv_l^j\over\Lambda^2}+
{\partial_m\partial_k\phi_l\partial_m\partial_k\phi_l\delta^{ij}
-2\partial_m\partial_i\phi_l\partial_m\partial_j\phi_l\over 8\pi G\,a^2\Lambda^2}+\mathcal{O}\!\left(1\over\Lambda^4\right)\ .\,\,\,\,\,\,\,\,\,\,
\eea
We see that $[\tau^{ij}]_\Lambda$ is sourced by short wavelength fluctuations plus higher derivative corrections.

Note that by taking the derivative $\partial_j$ this leading piece becomes
\bea
\partial_j[\tau^{ij}_s]_\Lambda=\partial_j\!\left[\sigma_s^{ij}\right]_\Lambda+\left[\rho_s\partial_i\phi_s\right]_\Lambda\ ,
\eea
with
\beq
[\rho_s\partial_i\phi_s]_\Lambda=\sum_{n\neq\bar{n}}m\,a^{-3}\partial_i\phi_{s,\bar{n}}({\bf x}_{n})W_\Lambda({\bf x}-{\bf x}_{n})-[\rho_l\partial_i\phi_s]_\Lambda\ , 
\label{shortgrav}\eeq
where the first term in (\ref{shortgrav}) is given by
\bea
&&\sum_{n\neq\bar{n}}m\,a^{-3}\partial_i\phi_{s,\bar{n}}({\bf x}_{n})W_\Lambda({\bf x}-{\bf x}_{n})\nonumber\\
&&=\sum_{n\neq\bar{n}}{m^2G\over a^4}{(x_n-x_{\bar{n}})^i \over |{\bf x}_n-{\bf x}_{\bar{n}}|^3}\left(
\mbox{Erfc}\left[{\Lambda|{\bf x}_n-{\bf x}_{\bar{n}}|\over\sqrt{2}}\right]+{4\pi|{\bf x}_n-{\bf x}_{\bar{n}}|\over\Lambda^2}W_\Lambda({\bf x}_n-{\bf x}_{\bar{n}})\right)W_\Lambda({\bf x}-{\bf x}_n)\ , \,\,\,\,\,\,\,\,\,\,\,\,\,\,\,
\eea
and the second term in (\ref{shortgrav}) can be expanded as
\beq
[\rho_l\partial_i\phi_s]_\Lambda=-{1\over 2\Lambda^2}\rho_l\partial_i\partial^2\phi_l+\ldots\ ,
\eeq
and this term should be included since it involves the background piece $\rho_b$, and so it includes a first order contribution.

\section{Expression for $\delta^{(3)}$ \label{app:delta3}}
The iterative solution for $\delta^{(3)}$ is given by
\bea
\label{eq:delta3}
&&\delta^{(3)}_l(\vec k,a)=\frac{1}{256\pi^6 D(a_0)^3} \\ \nonumber
&&\left[ 4{\cal D}_1(a) \left(\int d^3q\int d^3p \beta(\vec p,\vec q-\vec p)\delta s_1(\vec q-\vec p)\delta s_1(\vec p) \beta(\vkp,\vkkp)\delta s_1(\vkkp)\right.\right.\\ \nonumber
&&\  \ \qquad\left.\int d^3q\int d^3p \beta(\vec p,\vec k-\vec q -\vec p)\delta s_1(\vec k-\vec q-\vec p)\delta s_1(\vec p)\beta(\vec q,\vkkp)\delta s_1(\vkp)\right)+\\ \nonumber
&&2 {\cal D}_2(a)\int d^3q\int d^3p \beta(\vec p,\vec k-\vec q -\vec p)\delta s_1(\vec k-\vec q-\vec p)\delta s_1(\vec p)\alpha(\vec q,\vkkp)\delta s_1(\vkp)+\\ \nonumber
&&{\cal D}_3(a)\int d^3q\int d^3p \alpha(\vec p,\vec k-\vec q -\vec p)\delta s_1(\vec k-\vec q-\vec p)\delta s_1(\vec p)\alpha(\vec q,\vkkp)\delta s_1(\vkp)+\\ \nonumber
&&{\cal D}_4(a)\int d^3q\int d^3p  \alpha(\vec p,\vec q-\vec p)\delta s_1(\vec q-\vec p)\delta s_1(\vec p) \alpha(\vkp,\vkkp)\delta s_1(\vkkp)+
\\ \nonumber
&&2{\cal D}_5(a)\left(\int d^3q\int d^3p \alpha(\vec p,\vec q-\vec p)\delta s_1(\vec q-\vec p)\delta s_1(\vec p) \beta(\vkp,\vkkp)\delta s_1(\vkkp)\right.\\ \nonumber
&&\  \ \qquad\left.\int d^3q\int d^3p \alpha(\vec p,\vec k-\vec q -\vec p)\delta s_1(\vec k-\vec q-\vec p)\delta s_1(\vec p)\beta(\vec q,\vkkp)\delta s_1(\vkp)\right)+\\ \nonumber
&&\left.2{\cal D}_6(a) \int d^3q\int d^3p \beta(\vec p,\vec q-\vec p)\delta s_1(\vec q-\vec p)\delta s_1(\vec p) \alpha(\vkp,\vkkp)\delta s_1(\vkkp)\right]\ ,
\eea
where ${\cal D}_i$'s represent the result of the integration of the Green's functions and the other time-dependent coefficients. They are given by
\bea
&&{\cal D}_1=\int_0^{a}d\ta\;\ta^2\,G(a,\ta){\cal H}^2(\ta)D'(\ta)\int_0^{\ta}  d\hat a\; \hat a^2{\cal H}^2(\hat a)\d_{\ta}G(\ta,\hat a)D'(\hat a)^2\ ,\\ \nonumber
&&{\cal D}_2=3{\cal H}_0^2 \Omega_m\int_0^{a}d\ta\;\frac{G(a,\ta)}{\ta}\int_0^{\ta}  d\hat a\; \hat a^2{\cal H}^2(\hat a)D'(\hat a)^2  D(\ta)\,G(\ta,\hat a)+2 {\cal D}_1\ , \\ \nonumber
&&{\cal D}_3=\int_0^{a}d\ta\;\frac{G(a,\ta)}{\ta}\\ \nonumber
&&\ \int_0^{\ta}  d\hat a\; \frac{1}{\hat a}\left[3{\cal H}_0^2 \Omega_m D(\hat a)^2+2 \hat a^3{\cal H}^2(\hat a)D'(\hat a)^2\right]\left[3{\cal H}_0^2 \Omega_m D(\ta) G(\ta,\hat a)+2\ta^3{\cal H}^2(\ta)D'(\ta)\d_{\ta}G(\ta,\hat a)\right],\\ \nonumber
&&{\cal D}_4={\cal D}_3-4\int_0^{a} d\ta\; \ta^2\,G(a,\ta){\cal H}^2(\ta)D(\ta) D'(\ta)^2\ , \\ \nonumber
&&{\cal D}_5=3{\cal H}_0^2 \Omega_m \int_0^{a}d\ta\;\ta^2\,G(a,\ta)\;{\cal H}^2(\ta)D'(\ta)\int_0^{\ta} d\hat a\;D(\hat a)^2\frac{\d_{\ta}G(\ta,\hat a)}{\hat a}\\ \nonumber
&&\qquad\quad-2\int_0^{a} d\ta\; \ta^2\,G(a,\ta){\cal H}^2(\ta)D(\ta) D'(\ta)^2+2{\cal D}_1\ ,\\ \nonumber
&&{\cal D}_6={\cal D}_2+2\int_0^{a} d\ta\; \ta^2\,G(a,\ta){\cal H}^2(\ta)D(\ta) D'(\ta)^2\ .
\eea
Notice again the great simplification that occurs due to the fact that the growth factors and the Green's function do not depend on $k$, so that the time integrals and the momentum integrals decouple.

\section{${\cal O}(\delta_l^4)$ Power Spectrum with Approximate Treatment\label{app:approx_perturbation_theory}}

In the main part of the paper, we performed perturbation theory with our EFT in a rigorous and exact way. However, it is possible to perform an approximate treatment that simplifies quite a bit the formulas. In this way it is simpler to follow the derivation and we therefore present it here.

If the universe were to be EdS, then the solution for $\delta^{(n)}$ would be $\delta^{(n)}\propto a^n\simeq D(a)^n$. Thanks to this, all formulas simplify remarkably. Our universe is of course not of the EdS form,  because of the cosmological constant. But it is tempting to extend the results obtained in EdS to the ones in $\Lambda$CDM universe by replacing in the EdS formulas the EdS growth factor with the growth factor in $\Lambda$CDM. In this we obtain
\be
\langle\delta_l(\vec k,a_0)\delta_l(\vec q,a_0)\rangle_{\rm 1-loop}\simeq(2\pi)^3\delta^{(3)}(\vec k+\vec q)\left(\tilde P_{22}(k)+\tilde P_{13}(k)+\tilde P_{13,\;c^2_{\rm\,comb}}(k)\right)\ .
\ee
Here
$\tilde P_{22}$ and $\tilde P_{13}$ are the time-independent one loop contributions given by
\bea \nonumber
&&\tilde P_{22}(k)=\frac{k^3}{392 \pi ^2} \int_0^\infty dr \int_{-1}^1dx \frac{\left(-10 r x^2+3 r+7
   x\right)^2 }{\left(r^2-2 r x+1\right)^2} P_{11,l}(k r,\Lambda) P_{11,l}(k \sqrt{r^2-2 r
   x+1},\Lambda)\ , \\
&&\tilde P_{13}(k)=\frac{k^3 }{1008 \pi ^2} P_{11,l}(k,\Lambda)\\ \nonumber
&&\qquad \int_0^\infty dr  \left(\frac{3}{r^3} \left(r^2-1\right)^3 \left(7
   r^2+2\right) \log \left|\frac{r+1}{1-r}\right|-42 r^4+100
   r^2+\frac{12}{r^2}-158\right) P_{11,l}(k r,\Lambda) \ .
\eea
The counterterm contribution is given by
\be
\tilde P_{13,\;c^2_{\rm\,comb}}(k)=-\frac{2}{9}\frac{c_{\rm\,comb}^2(a_0) D(a_0)^2}{ {\cal H}_0^2D'(a_0)^2a_0^2} k^2 P_{11,l}(k,\Lambda)\ ,
\ee
where the time dependence of $c^2_{\rm\,comb}$ can be inferred in perturbation theory from (\ref{eq:parameters}) to be
\be
c^2_{\rm\,comb}(a)=c^2_{\rm\,comb}(a_0) \frac{{\cal H}^2(a) D'(a)^2 a^2}{{\cal H}_0^2 D'(a_0)^2a_0^2}\ .
\ee
Notice that, within this approximation, the $k$ dependence and the time dependence of $c^2_{\rm\,comb} \delta_l^{(1)}$ is the same as the one of the source of $\delta^{(3)}_l$ in the high $k$ limit.

\begin{figure}[h!]
\begin{center}
\includegraphics[width=8cm]{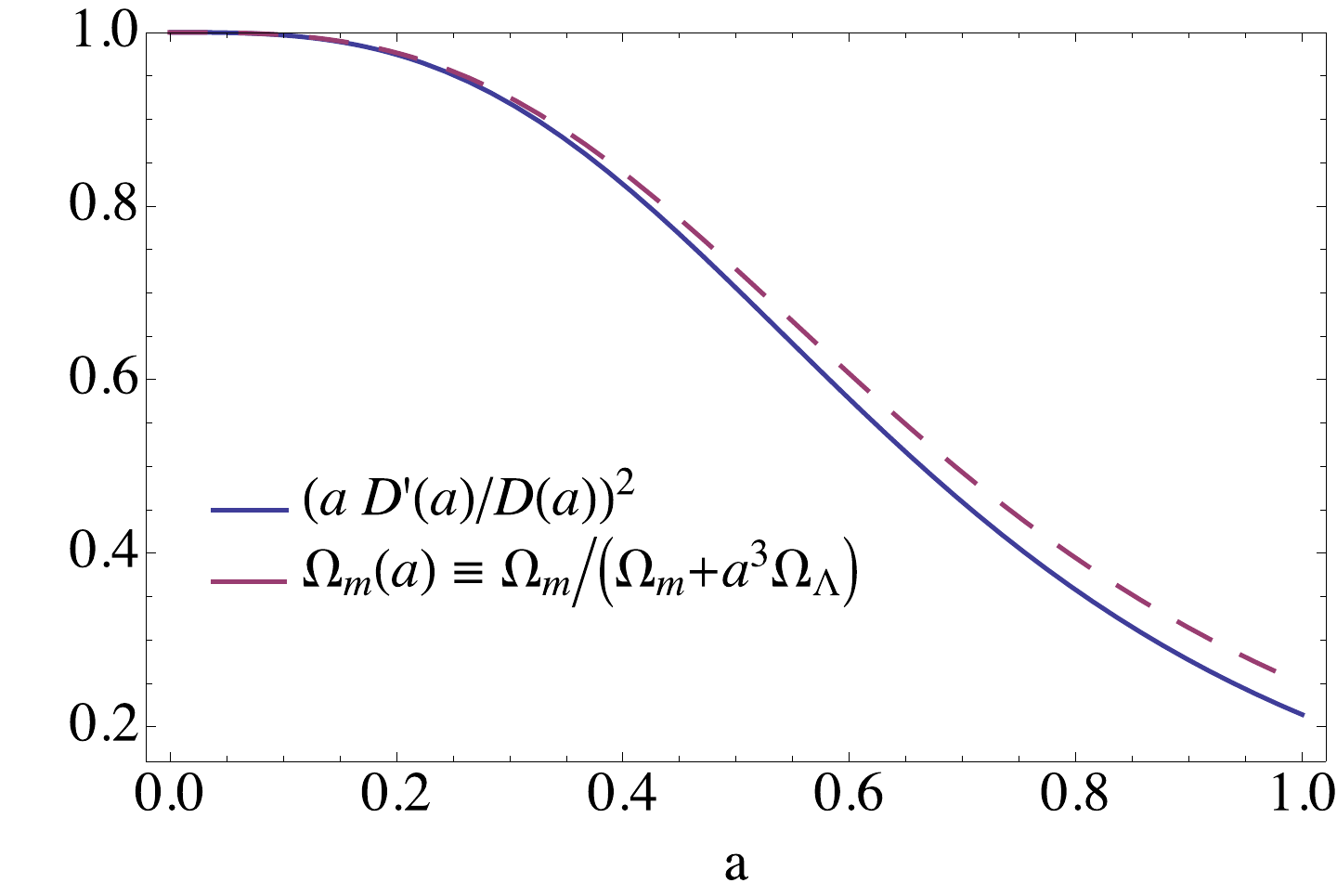}
\includegraphics[width=8cm]{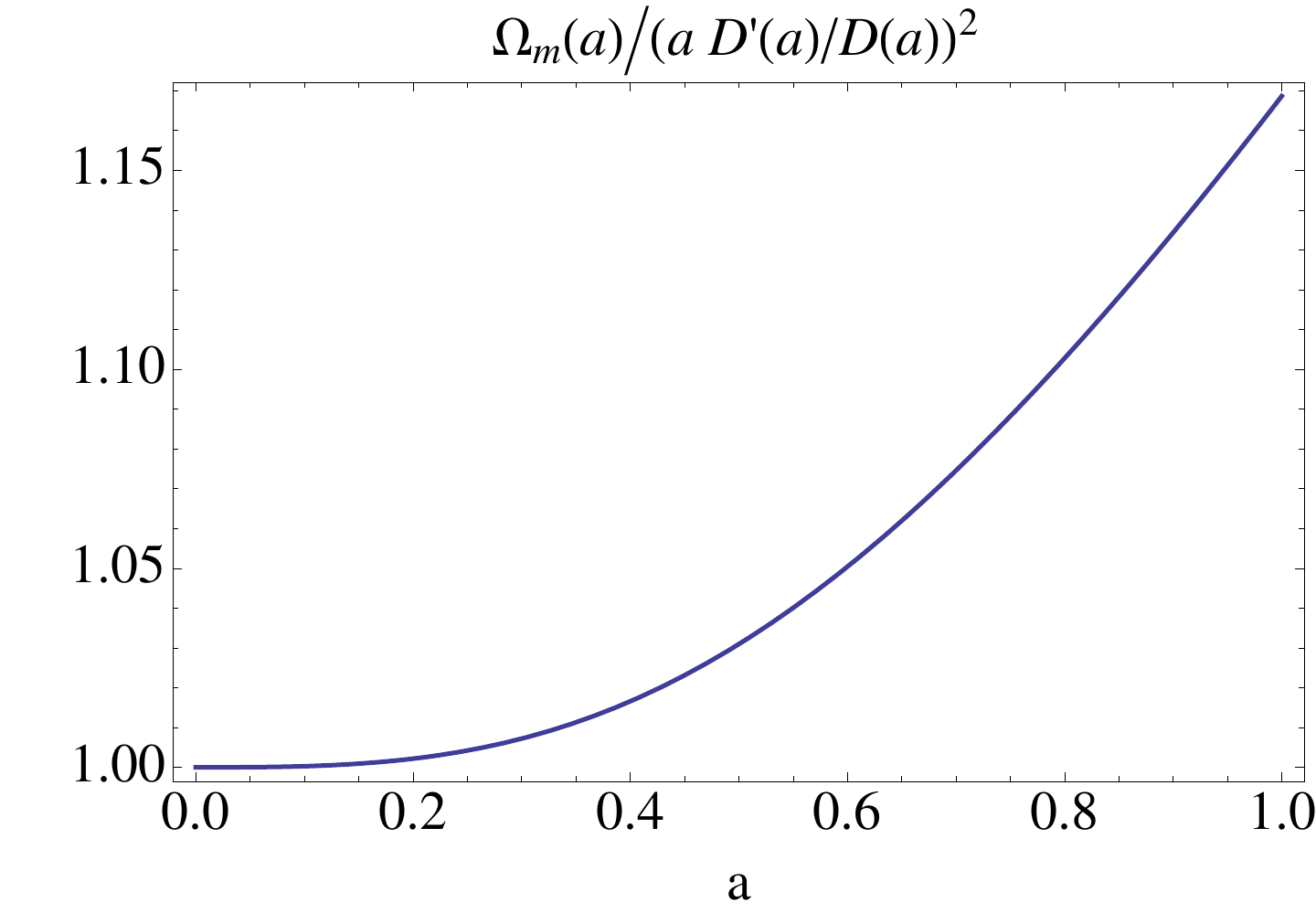}
\caption{\label{GrowthFuncComparison} {\small \it Plot of the ratio $\Omega_m(a)/(\d\log D/\d\log a)^2$. This being equal to one would justify the approximations done in this appendix. We see that at late times the difference is quite large. Since this is a correction to the one-loop term,  this error can be acceptable for a one-loop calculation. On the contrary the approximation can become more harmful if one goes to higher loops. The actual result on the power spectrum of the approximation is even better than what shown in the plot, as more of the clustering happens before dark energy domination.}}
\end{center}
\end{figure}

This approximation is quite a good numerical approximation, and here below in Fig.~\ref{fig:time_comparison} we present results of comparisons for $P_{22}$ and $P_{13}$, where we see that the disagreement is at percent level. This ratio is plotted in Fig.~\ref{GrowthFuncComparison}. Notice that at $k\sim 0.24$ Mpc$^{-1}$ where the non-linear corrections are of order of a few ten percents, a percent error in the loop calculation leads to an error that is dangerously close to one percent. Therefore, for percent precision, using this approximate treatment at the high $k$'s that we can reach with the EFT corresponds to pushing the boundaries of safety. It is further important to stress that this is not a parametrically good approximation in $\delta_l$, as is our loop expansion. As noted in~\cite{martel:1991,Bernardeau:1993qu,Scoccimarro:1997st}, this is an expansion in the smallness of the ratio $\Omega_m(a)/(\d\log D/\d\log a)^2$. A good fit is $(\d\log D/\d\log a)^2\sim \Omega(a)^{1.2}$, explaining the possibility to make this approximation. 
Luckily, we find it not so hard to implement directly the correct perturbation theory, which is the one we present in this paper.

\begin{figure}[h!]
\begin{center}
\includegraphics[width=8cm]{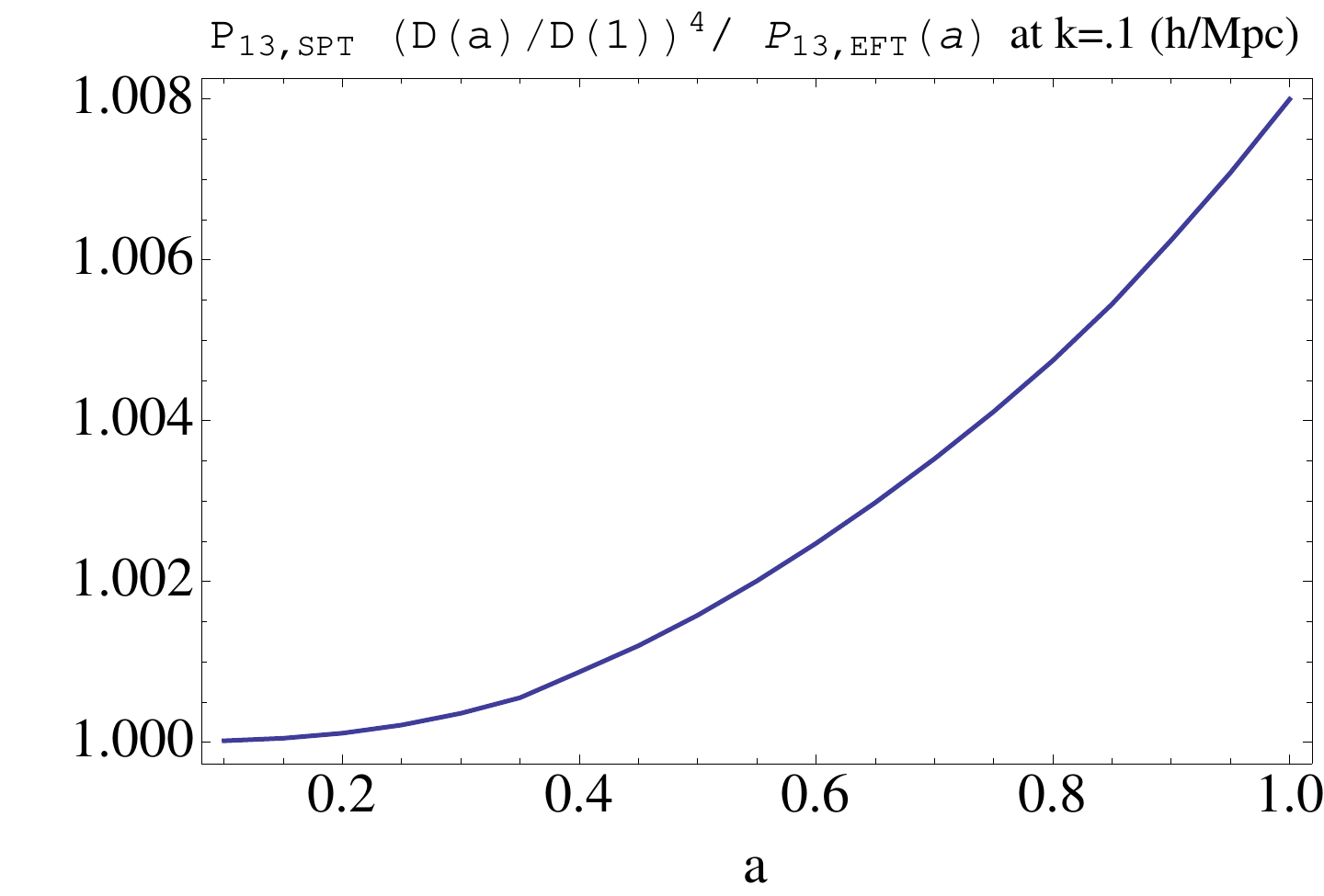}
\includegraphics[width=8cm]{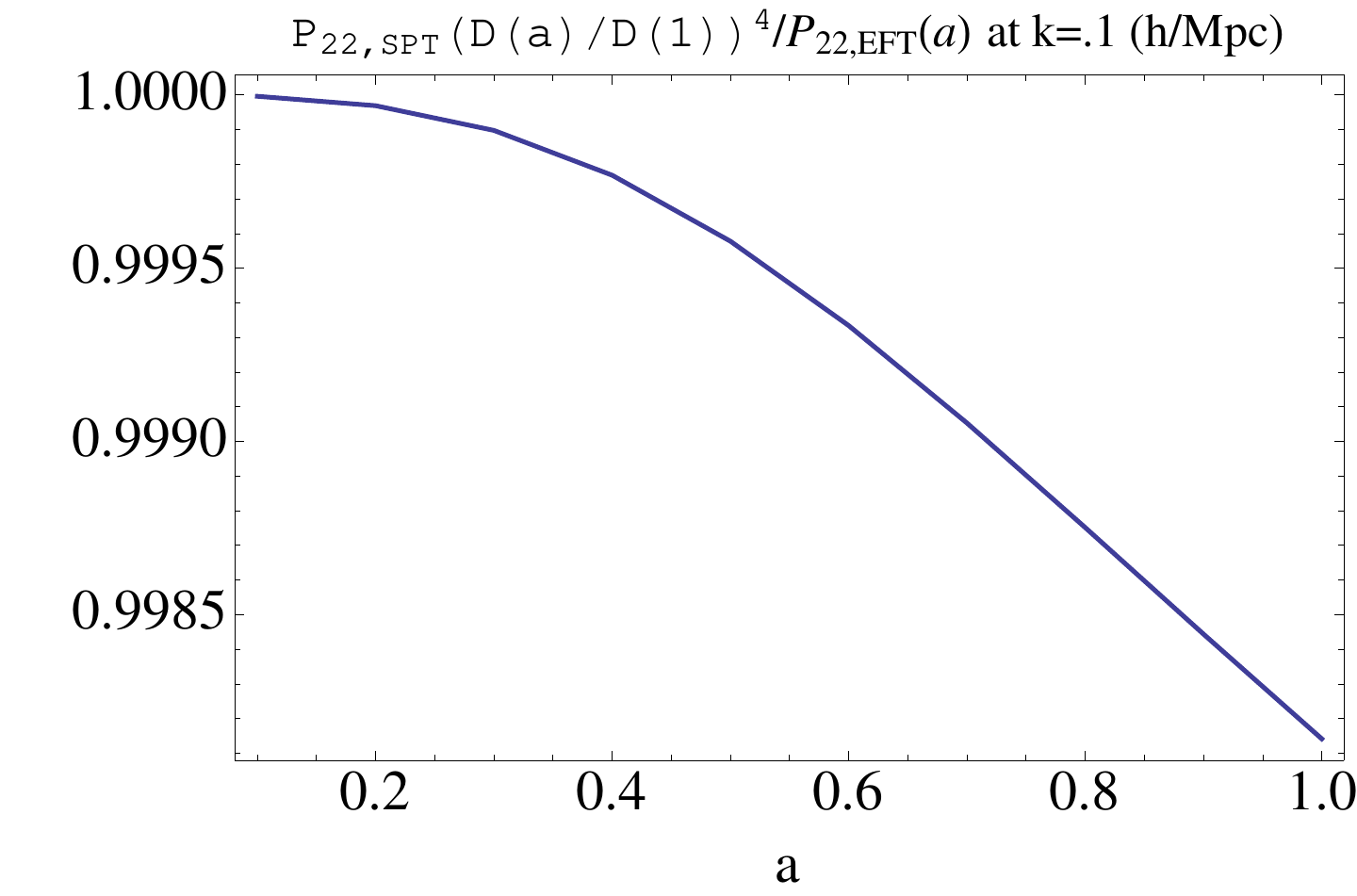}
\caption{\label{fig:time_comparison} \small\it We plot the ratio as a function of time of $P_{13}$ and $P_{22}$ as obtained using the rigorous  time dependence and the approximate one. The ratio goes to 1 at early times when dark energy is irrelevant and the approximate treatment becomes exact. The results at redshift zero agree well, at percent level. }
\end{center}
\end{figure}

\section{Measuring in $N$-body simulations\label{app:simulations}}

\def\xmxn{ { \myR - \vec{r}_n }}
\def\dxmxn{(  { \myR - \vec{r}_n })^2}

\subsection{Efficient calculation}
\label{App:MeasurementDetails}

We break the calculation into three parts:
 the calculation of {\it primary fields}, 
 the calculation of {\it secondary fields}, and
 the calculation of correlations. 
Primary fields are dependent solely upon the positions or velocities of the 
simulation particles, and secondary fields are dependent upon the primary fields,
and we care about correlations between particular secondary fields.  

The immediate goal is to calculate the following expressions:
\begin{align}
c_s^2(r)&= \frac{ P_{A\Theta} \partial^2 P_{\delta \Theta} - P_{A \delta} \partial^2 P_{\Theta \Theta}}{ ( \partial^2 P_{\delta \Theta})^2  - \partial^2 P_{\delta \delta} \partial^2 P_{\Theta \Theta}}\\  
c_v^2(r)&= \frac{ P_{A\delta} \partial^2 P_{\delta \Theta} - P_{A \Theta} \partial^2 P_{\delta \delta}}{ ( \partial^2 P_{\delta \Theta})^2  - \partial^2 P_{\delta \delta} \partial^2 P_{\Theta \Theta}  }
\end{align}
based upon the two-point correlation functions: $P_{IJ}(r)$.   These two point correlation functions can be seen as the expectation value of the product of field $I$ with field $J$, but with $I$ evaluated at all points a distance $r$ from all points $r'$:
\begin{equation}
P_{IJ}(r)\equiv \langle  I( \vec{r}+\vec{r}\,') J(\vec{r}\,') \rangle_{\Omega(r',r),r'} 
=\lim_{R \rightarrow \infty} \frac{3}{16 \pi^2 R^3}  \int_0^R  \, dr' \,  d\Omega' \, d\Omega \, r'^2 \,  I(\vec{r} + \vec{r}\,') \, J(\vec r\,').
\end{equation}  
In the quantities being calculated the overall normalization cancels.   The cost of evaluating each of these fields at all positions is prohibitive, so we approximate these correlations by measuring some large $N$ number of pairs of points (pseudo)-randomly chosen but at fixed $r$:
\begin{equation}
\hat{P}_{I J}(r)\approx  N^{-1} \!\!\!\!\!\!\!\!  \sum_{\{\vec r_a\,',\vec r_b\,'\}  \in N_{\rm set}(r)}  \!\!\!\!\!\!\!\!  I( \vec r_a\,') ~ J (\vec r_b\,')
\end{equation}
where $N_{\rm set}(r)$ is a set of $N$ pairs of points randomly selected from the space to be separated by a distance $r$.

The following are the specific correlation functions used:
\begin{align}
P_{A\delta}(r)&= \langle  A_s( \vec{r}+\vec{r}\,') \delta_l(\vec{r}\,') \rangle_{\Omega(r',r),r'}\\
P_{A\Theta}(r)&= \langle  A_s( \vec{r}+\vec{r}\,' ) \Theta_l(\vec{r}\,')  \rangle_{\Omega(r',r),r'}  \\
P_{\delta\delta}(r)&= \langle \delta_l( \vec{r}+\vec{r}\,' ) \delta_l(\vec{r}\,')  \rangle_{\Omega(r',r),r'}\\
P_{\delta\Theta}(r)&=  \langle  \delta_l( \vec{r}+\vec{r}\,' ) \Theta_l(\vec{r}\,')  \rangle_{\Omega(r',r),r'}\\
P_{\Theta\Theta}(r)&= \langle  \Theta_l( \vec{r}+\vec{r}\,' ) \Theta_l(\vec{r}\,') \rangle_{\Omega(r',r),r'} \\
\partial^2 P_{\delta\delta}(r)&= \langle \partial^2 \delta_l( \vec{r}+\vec{r}\,' ) \, \delta_l(\vec{r}\,')  \rangle_{\Omega(r',r),r'}\\
\partial^2 P_{\delta\Theta}(r)&=  \langle  \partial^2 \delta_l( \vec{r}+\vec{r}\,' ) \, \Theta_l(\vec{r}\,')  \rangle_{\Omega(r',r),r'}\\
\partial^2 P_{\Theta\Theta}(r)&= \langle  \partial^2 \Theta_l( \vec{r}+\vec{r}\,' ) \, \Theta_l(\vec{r}\,') \rangle_{\Omega(r',r),r'} 
\end{align}
These are functions of the gravitational short-mode field $A_s$, the over density $\delta$, and the velocity divergence $\Theta$, and relevant spatial derivatives. These fields are to be calculated from the observed (or simulated) positions and velocities of point-sources at a fixed moment in time (redshift=0 initially).  We  smear the observation of the position of these sources with a gaussian, introducing a (soft) ultraviolet cutoff.  

To calculate $c_s^2$ and $c_v^2$ in terms of $A_s, \delta, \Theta, \partial^2 \delta,\partial^2 \Theta$ we use a number of secondary fields.  While it would be possible to numerically estimate the relative necessary spatial derivatives, it is simple enough to explicitly carry the operations out analytically, and treat them as independent secondary fields, avoiding the introduction of specious numerical error.   
\begin{align}
[ \delta] (\myR) &= [ \rho] (\myR)/\rho_b -1\, ,\\
[ \Theta] (\myR) &=-\frac{1}{H a} \sum_{i=1}^{3}  {[\partial_i  v_i ](\myR) }\, \\
[ A_s] (\myR) &=\frac{1}{\rho_b}  \sum_{i=1}^{3} {\left([\partial_i \partial_i \phi_s](\myR) + \sum_{j=1}^3 [\partial_i \partial_j \kappa_{ij}](\myR) \right)}\,,\\
[ \partial^2 \delta] (\myR) &=\sum_{i=1}^{3} [ \partial_i^2 \rho] (\myR)/\rho_b \, ,\\
[ \partial^2 \Theta] (\myR) &=-\frac{1}{H a} \sum_{i=1, j=1}^{3}  {[\partial_i^2 \partial_j   v_j ](\myR) }\, \\
[\partial_i \partial_j \phi_s](\myR) &= [ \partial_i [\rho_m \partial_j \phi]_{\Lambda s}) ](\myR) + \frac{4 \pi G a^2}{2 \Lambda^2}  \left([\partial_i \rho](\myR)  [\partial_j \rho](\myR) +[\rho](\myR) [\partial_i \partial_j \rho](\myR) \right)
\end{align}
\begin{align}
[\partial_i \partial_j \kappa_{ij}](\myR) &= [\partial_i \partial_j \sigma_{ij}] (\myR) -
[\partial_j \pi_i](\myR)\, [\partial_i v_j](\myR) - 
[\partial_i \pi_i](\myR) [\partial_j v_j](\myR)\\
&~-[\pi_i] (\myR) \, [\partial_i \partial_j v_j](\myR) -
[\partial_i \partial_j \pi_i] (\myR) \, [v_j](\myR) \nonumber \\
[v_i](\myR)&=[\pi_i](\myR)/[\rho](\myR)\\
[\partial_i v_j](\myR)&=[\partial_i \pi_j](\myR)/[\rho](\myR)-[\pi_j](\myR) [\partial_i \rho](\myR)/([\rho](\myR))^2\\\
[\partial_i \partial_j v_k](\myR)&=[\partial_i \partial_j \pi_k](\myR)/[\rho](\myR)-
[\partial_i \pi_k](\myR) [\partial_j \rho] /([\rho](\myR))^2-
[\partial_j \pi_k](\myR) [\partial_i \rho] /([\rho](\myR))^2\\
&-[\pi_k](\myR) [\partial_i \partial_j \rho](\myR)/([\rho](\myR))^2
+2[\pi_k](\myR) [\partial_i \rho](\myR) [\partial_j \rho](\myR)/([\rho](\myR))^3\nonumber\\
[\partial_i^2 \partial_j v_j](\myR)&= ([\rho](\myR))^{-4} \Big\{
            -6 \,  ([\partial_i \rho](\myR))^2  \, [\partial_j \rho](\myR) \, [\pi_j](\myR) \\
            &~+ 2 \, \Big(
                 ([\partial_i \rho](\myR))^2  \, [\partial_j \pi_j](\myR) +
                  [\partial_i \partial_i \rho](\myR) \, [\partial_j \rho](\myR) \, [\pi_j](\myR) \nonumber\\
             &~ +
                  2\,[\rho](\myR) \, [\partial_i \rho](\myR) \, ( [\partial_i \pi_j](\myR) \, [\partial_j \rho](\myR) + [\partial_i \partial_j \rho](\myR) \, [\pi_j](\myR) )
              \Big)  \nonumber\\
              &~ -\, \, ( [\rho](\myR))^2 \, \Big(
                2 \, [\partial_i \partial_j \rho](\myR) \, [\partial_i \pi_j](\myR) +
                  2 \, [\partial_i \partial_j\pi_j](\myR) \, ([\partial_i \rho](\myR))  \nonumber\\
                  &~+
                  [\partial_i \partial_i \rho](\myR) \, [\partial_j \pi_j](\myR)+[\partial_i \partial_i\pi_j](\myR) \, [\partial_j \rho](\myR) +
                  [\partial_i^2 \partial_j\rho](\myR) \, [\pi_j](\myR)
              \Big) \nonumber\\
              &~+ ( [\rho](\myR))^3\, [\partial_i \partial_i \partial_j \pi_j](\myR)  
        \Big\}\nonumber
\end{align}
These are, in turn, defined in terms of the following primary fields
  
\begin{align}
\left [ \rho \right ](\myR) &=m \Lambda^3 / (a^3  (2 \pi)^{3/2})  \sum_{n \in \eta(\myR)}   W( \myR -\vec{r}_n)  \\
\left [ \partial_i  \rho \right ](\myR) &=  m \Lambda^3 / (a^3  (2 \pi)^{3/2})  \sum_{n \in \eta(\myR)}    \partial_i W( \myR - \vec{r}_n) \\
\left [\partial_i \partial_j \rho \right](\myR) &= m \Lambda^3 / (a^3  (2 \pi)^{3/2})  \sum_{n \in \eta(\myR)}    \partial_i  \partial_j W( \myR - \vec{r}_n)
 \end{align}
 \begin{align}
\left [(\partial_i)^2 \partial_j \rho \right](\myR) &=   m \Lambda^3 / (a^3  (2 \pi)^{3/2})    \sum_{n \in \eta(\myR)} (\partial_i)^2  \partial_j W( \myR - \vec{r}_n)\\ 
\left[ \pi_i \right](\myR) &= m  \Lambda^3 / (a^3  (2 \pi)^{3/2})  \sum_{n \in \eta(\myR)}   W( \myR -\vec{r}_n)
 (\vec{v}_n)_i  \\
\left[ \partial_i \pi_j \right](\myR) &= m  \Lambda^3 / (a^3  (2 \pi)^{3/2})  \sum_{n \in \eta(\myR)}   \partial_i W( \myR -\vec{r}_n)  (\vec{v}_n)_j \\
\left[ \partial_i \partial_j  \pi_k \right](\myR) &= m  \Lambda^3 / (a^3  (2 \pi)^{3/2})  \sum_{n \in \eta(\myR)}  
 \partial_i  \partial_j W( \myR -\vec{r}_n)  (\vec{v}_n)_k \\
\left[ ( \partial_i )^2 \partial_j  \pi_k \right](\myR) &= m  \Lambda^3 / (a^3  (2 \pi)^{3/2})  \sum_{n \in \eta(\myR)}   (\partial_i)^2  \partial_j W( \myR -\vec{r}_n)  (\vec{v}_n)_k \\
\left[ \partial_i \partial_j \sigma_{ij} \right ](\myR) &=  m  \Lambda^3 / (a^3  (2 \pi)^{3/2})   \sum_{n \in \eta(\myR)} 
 \partial_i  \partial_j W( \myR -\vec{r}_n)  (\vec{v}_n)_i (\vec{v}_n)_j \\
\left[ \partial_i [ \rho_m   \partial_j \phi]_{\Lambda s} \right](\myR) &=
m^2 G / a^4  (\Lambda / \sqrt{2 \pi} )^3  \sum_{n \in \eta(\myR)} (\partial_i W(\xmxn) )   (\xi_n)_j
\end{align}
where $\mu \equiv r_{\rm max}$ and $\eta(\myR)=\{n ~ {\rm s.t.}~ \left| \xmxn\right| < r_{\rm max} \}$, and we have introduced the following functions for notational convenience:
\begin{align}
W(r) &\equiv  \exp \left(-\frac{\Lambda^2}{2} r^2\right)\\\
\partial_i W(r) &\equiv -\Lambda^2 r_i W(r) \\ 
\partial_i \partial_j W(r) &\equiv \Lambda^2 W(r) ( \Lambda^2 r_i r_j -\delta^{ij}_{\rm k} )\\
(\partial_i)^2 \partial_j W(r) &\equiv \Lambda^2 \left( \partial_i W(r) \left( \Lambda^2 r_i r_j -\delta^{ij}_{\rm k} \right )+ W(r)  \Lambda^2 ( { \mathbf 1}_i \,  r_j + \delta^{ij}_{\rm k} r_i) )\right )  \\
(\xi_n)_j &\equiv \sum_{m \in \eta_n} (\vec{r}_n-\vec{r}_m)_j \, \Gamma(\vec{r}_n -\vec{r}_m) \, \Bigg( \frac{1}{ |\vec{r}_n - \vec{r}_m |} {\rm erfc}  (\Lambda |\vec{r}_n-\vec{r}_m| /\sqrt{2})\nonumber\\
&~~~~+\sqrt{\frac{2}{\pi}}\,  \Lambda \,  W(\vec{r}_n-\vec{r}_m) \Bigg)\, ,
\end{align}
where $\eta_n = \{ m~{\rm s.t.}~ |r_m - r_n | < r_{\rm max} \}$.  We set $r_{\rm max} \equiv 7 \Lambda$.  It is efficient to precalculate and store $(\xi_n)_j$ for a given downsample of particles and $\Lambda$.    Given that each simulation particle has a limited radius of influence $r_{\rm max}$, a variety of parallelization strategies are available.    We have presented these fields, in detail, to emphasize that all such secondary fields rely on a relatively small number of primary fields, and it is only the primary fields which need concern themselves with the explicit number of particles down sampled from the simulation.

\subsection{Stability of Measurement and Region Selection}
\label{App:Correlation plots}
\begin{figure}
    \centering
        \includegraphics[width=0.49\textwidth]{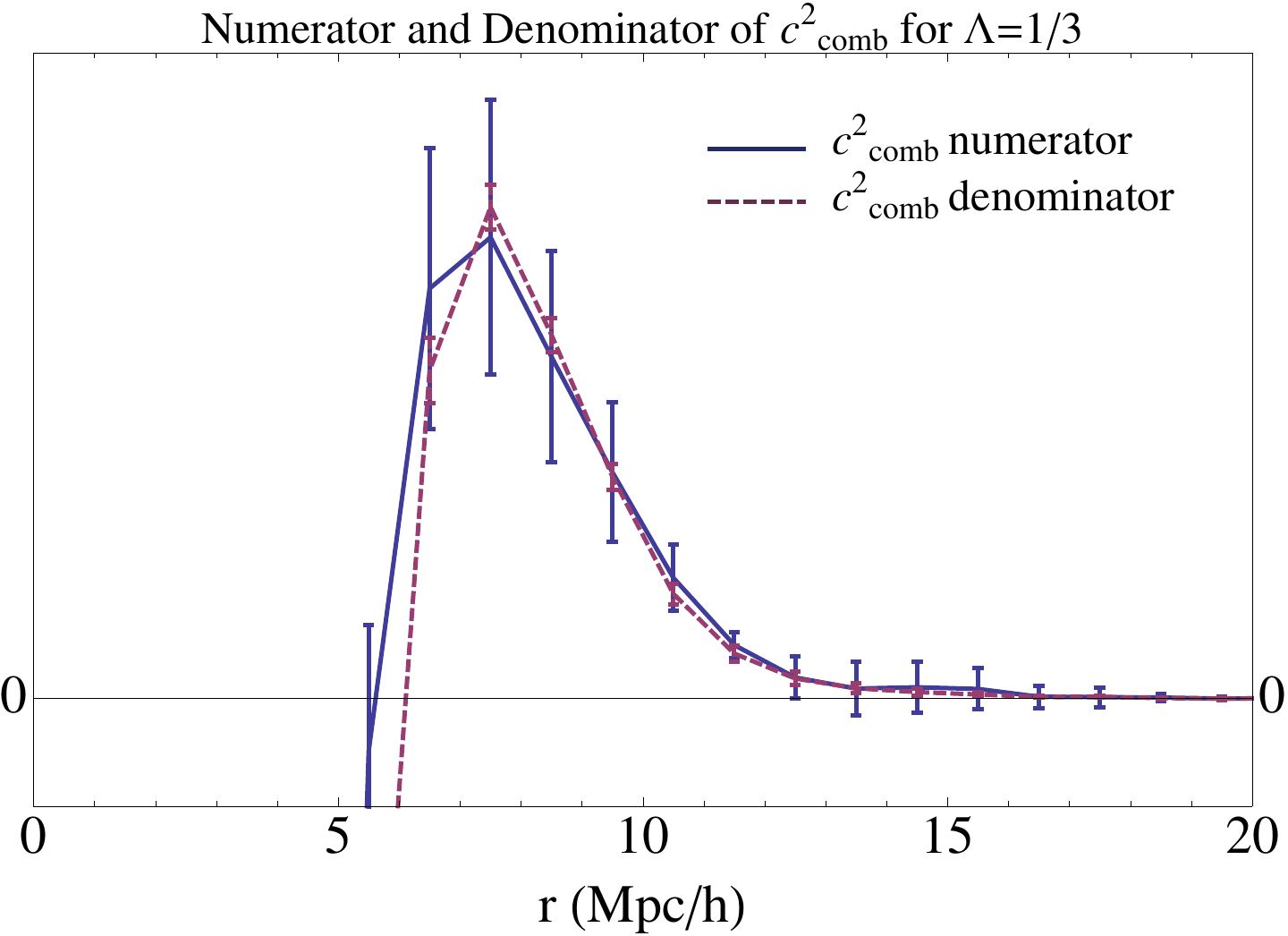}
         \includegraphics[width=0.49\textwidth]{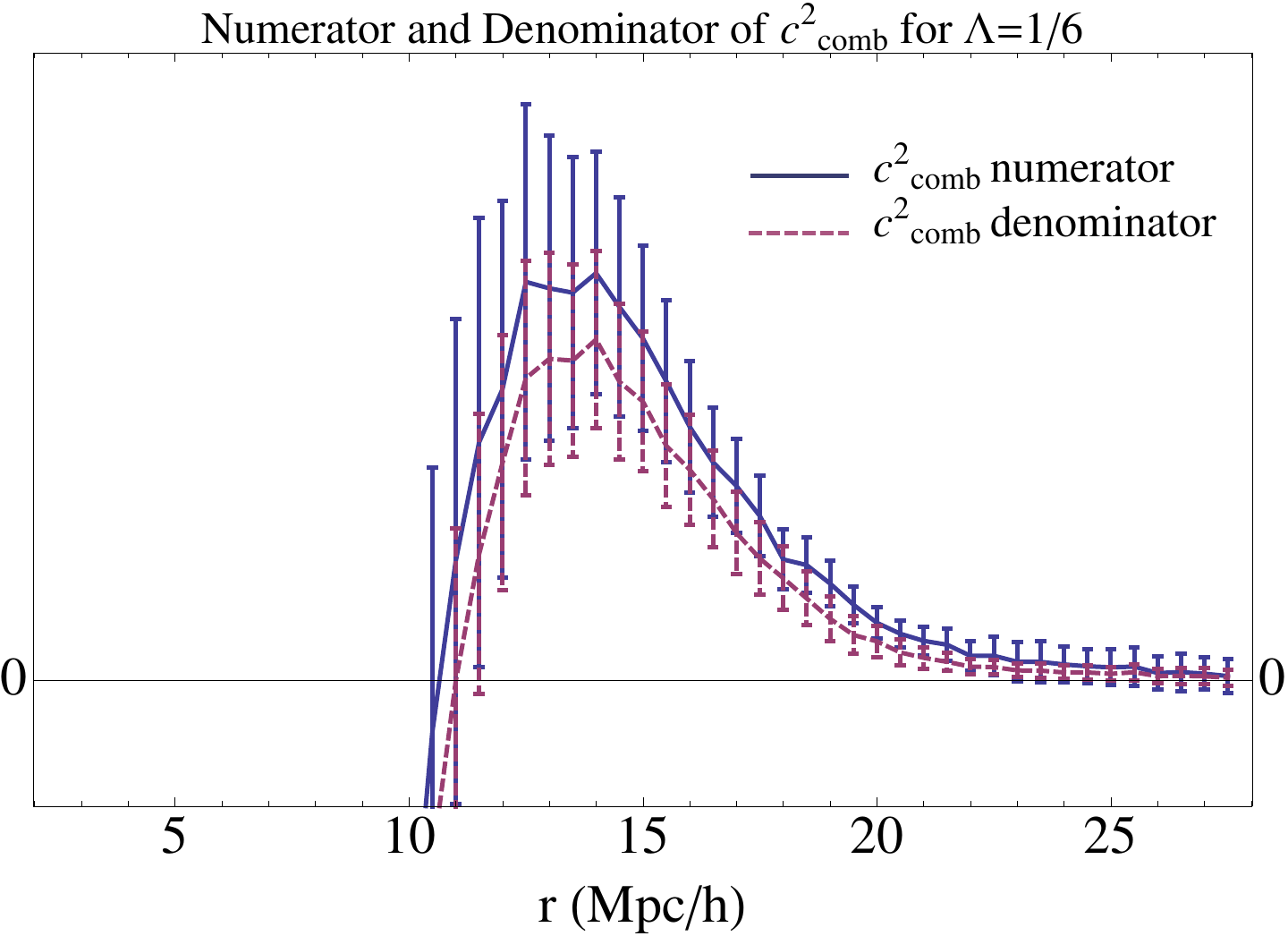}
    \caption{\sl  The numerator and denominators of $\csc$ as measured with smoothing parameter $\Lambda=1/3$ (left) and $\Lambda=1/6$ (right), scaled to similar heights.  This allows us to choose a convenient region of measurement to avoid zero over zero contamination.  Precision calculations in the future should extend measurements farther into the IR.   }
    \label{comparisons}
\end{figure}

The fluid parameter of interest $\csc$ is calculated as a ratio of polynomial functions of correlations.  In \fig{comparisons} we plot the numerator and denominator of these ratios for $\Lambda=1/3\,(h/Mpc)$ and $\Lambda=1/6\,(h/Mpc)$.  It is worth noting strong confirmation of the effective field theory description is how well the numerator and denominator of the $\csc$ ratio tracks each other.  With these sorts of statistical measurements, one should be careful of small fluctuations causing misleading signal near zero over zero regions.  For the measurements  described in the paper we select a region where the denominator is  $3 \sigma$ above zero.

\section{Renormalizing at finite $\Lambda$\label{app:higher-deriv}}

\begin{figure}[h!]
\begin{center}
\includegraphics[width=12cm]{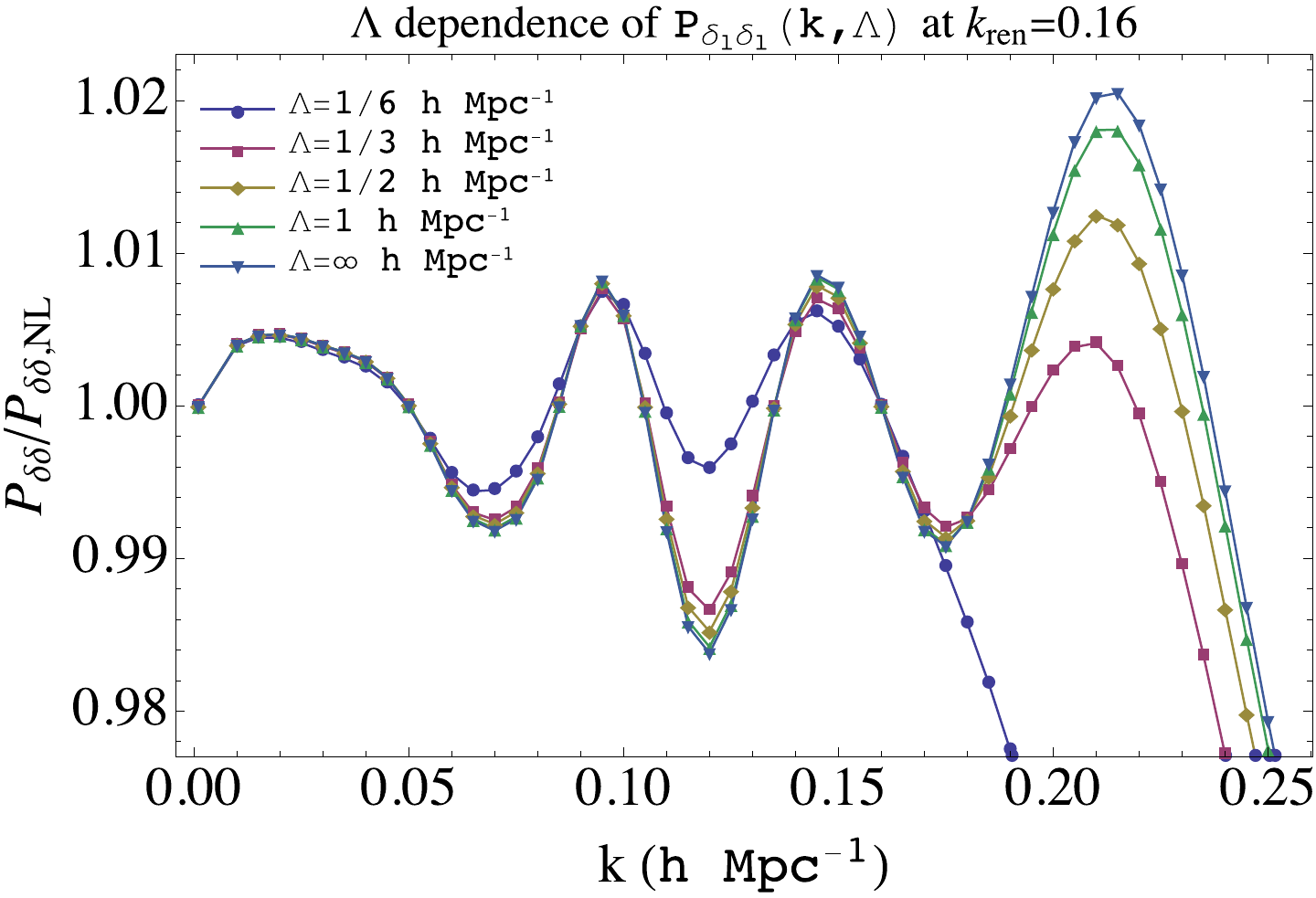}
\caption{\label{fig:finite_lambda} \small\it Prediction of the non-linear power spectrum without the addition of higher derivative terms, as we send $\Lambda\to\infty$, normalized to the non-linear power spectrum. We see that if we keep $\Lambda$ finite, non-included higher derivative terms that scale as powers of $k/\Lambda$ are important. Indeed the results improves as $\Lambda=\infty$, which is the correct procedure.}
\end{center}
\end{figure}

\begin{figure}[h!]
\begin{center}
\includegraphics[width=12cm]{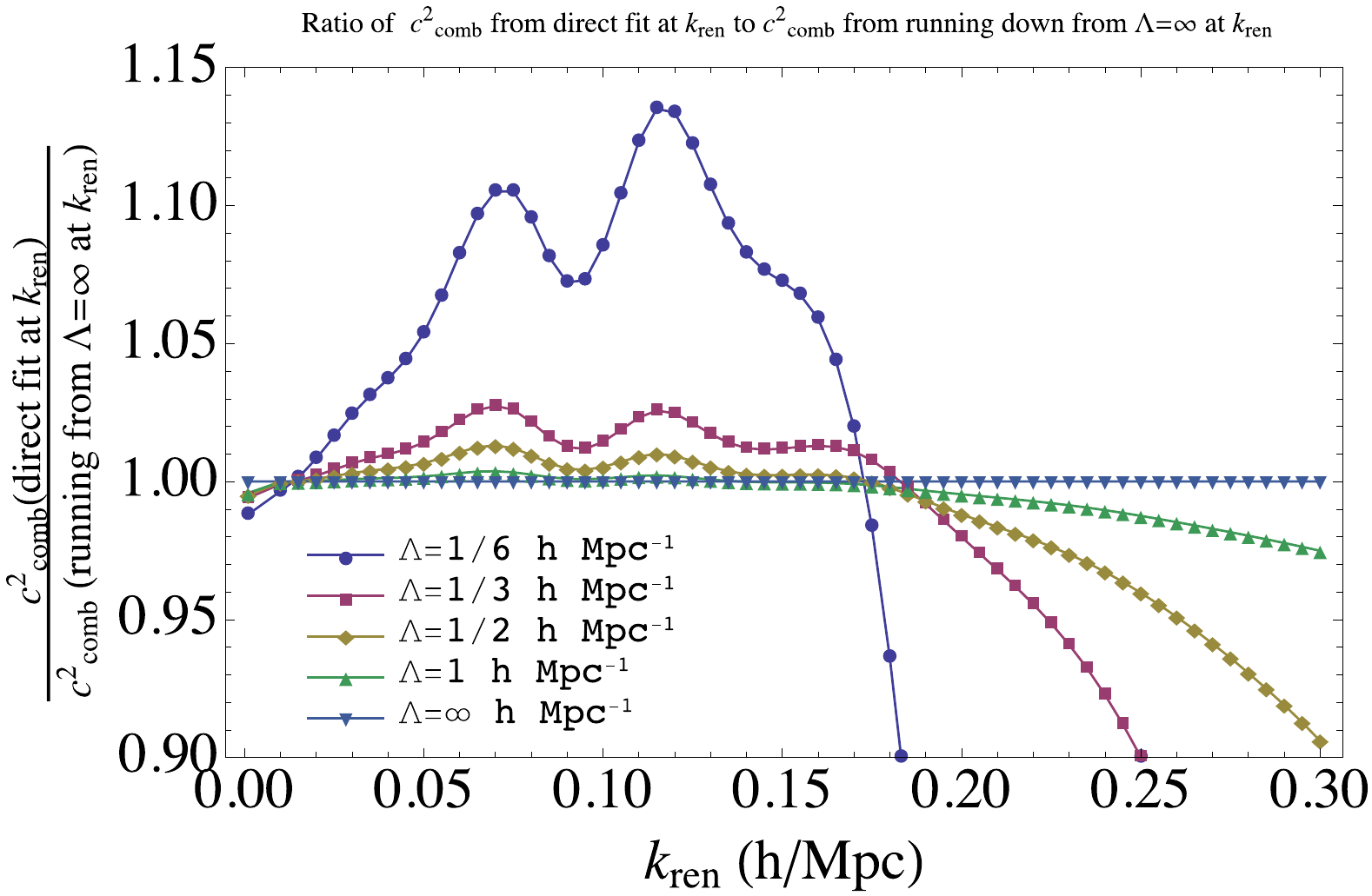}
\caption{\label{fig:cscombwrong} \small\it  Ratio of the value of $c_{\rm comb}^2(\Lambda,k_{\rm ren})$ as obtained from running from $c_{\rm comb}^2(\Lambda=\infty,k_{\rm ren})$ versus the one obtained by fitting directly the result of the EFT at finite $\Lambda$ without the inclusion of higher derivative terms in $k/\Lambda$. At low $k_{\rm ren}$, the error is particularly pronounced for $\Lambda=1/6$, as in that case $k_{\rm ren}/\Lambda$ is not very small.  Inclusion of higher derivative terms reduces the mismatch to few percent.}
\end{center}
\end{figure}

The procedure as outlined in the main text involves renormalization for $\Lambda=\infty$ at some chosen $k_{\rm ren}$ by fitting $P_{\delta\delta}^{\rm 1-loop}$ for  $\csc(\Lambda)$ against observation at that particular $k_{\rm ren}$.  The power spectrum at all  other $k$ become predictions of the EFT. 
We could imagine to perform the same procedure at finite $\Lambda$. In this case, however, higher derivative terms suppressed by powers of $k/\Lambda$ should be included. These terms do indeed vanish as $\Lambda\to \infty$, but at finite $\Lambda$  and $k$ they are not negligible. In fact, as shown in Fig.~\ref{fig:finite_lambda}, without the addition of higher derivative terms, the power spectrum deviates from the $\Lambda=\infty$ as $k\to\Lambda$. One can indeed check that the values of $c^2_{\rm comb}$ that is obtained by fitting in this way is off with respect to the correct value as obtained from running down to finite $\Lambda$ the value of $c^2_{\rm comb}$ at $\Lambda=\infty$ by about 15\%, depending on the cutoff used, see Fig.~\ref{fig:cscombwrong}. Instead, by allowing for higher derivative terms, the correct value of $c^2_{\rm comb}$ is derived. 

\end{appendix}
\clearpage

\begingroup\raggedright\endgroup

\end{document}